\documentclass[aps,graphics,a4paper,10pt,twocolumn,superscriptaddress,nofootinbib]{revtex4-2}

\usepackage[utf8]{inputenc}

\usepackage[margin=25mm]{geometry}

\usepackage{amsmath}
\usepackage{braket}
\usepackage{hyperref}
\usepackage{textcomp}
\usepackage{xcolor}

\usepackage{tikz}
\usepackage{pgfplots}
\usepackage{tikzscale}
\usepackage{bbm}
\usepackage{commath}

\usepackage{environ}
\usepackage[T1]{fontenc}
\usepackage{tgtermes}

 \usepackage{graphicx}
\usepackage{amsmath}
\usepackage{amsfonts}
\usepackage{amssymb}
\usepackage{soul,color}
\usepackage{dsfont}
\hypersetup{colorlinks=true, citecolor=blue, urlcolor=blue, linkcolor=blue}
\usepackage{amstext}
\usepackage[caption=false]{subfig}
\usepackage{epstopdf} %converting to PDF
\usepackage{mathtools} %text over arrows
\usepackage{physics}
\usepackage{float}

\usepackage{bm}
\usepackage{braket}
\usepackage{placeins}

\usepackage{diagbox}%slash box in table
\usepackage{hhline}

\usetikzlibrary{matrix}

\makeatletter
\newsavebox{\measure@tikzpicture}
\NewEnviron{scaletikzpicturetowidth}[1]{%
  \def\tikz@width{#1}%
  \begin{lrbox}{\measure@tikzpicture}%
  \BODY
  \end{lrbox}%
  \pgfmathparse{#1/\wd\measure@tikzpicture}%
  \BODY
}
\makeatother

\usepackage{graphicx,kantlipsum,setspace}
\usepackage{caption}
\usepackage{soul}
\usepackage[title,page]{appendix}

\newcommand{\avg}[1]{\langle #1\rangle}
\newcommand{\be}{\begin{equation}\begin{aligned}} 
\newcommand{\ee}{\end{aligned}\end{equation}}
\newcommand{\beN}{\begin{equation*}\begin{aligned}} 
\newcommand{\eeN}{\end{aligned}\end{equation*}}
\newcommand{\ba}{\begin{array}}
\newcommand{\ea}{\end{array}}
\newcommand{\bqa}{\begin{eqnarray}}
\newcommand{\eqa}{\end{eqnarray}}

\newcommand{\eq}[1]{Eq.~\eqref{eq:#1}}

\newcommand{\secname}[1]{Sec.~\ref{sec:#1}}
\newcommand{\apdxname}[1]{Appendix.~\ref{apdx:#1}}
\newcommand{\Figname}[1]{Figure~\ref{fig:#1}}
\newcommand{\figname}[1]{Fig.~\ref{fig:#1}}

\newcommand{\brkts}[1]{\left( #1 \right)}

\newcommand{\RN}[1]{\textup{\uppercase\expandafter{\romannumeral#1}}}

\renewcommand{\cal}{\mathcal}

\newcommand{\ha}{\hat{a}}
\newcommand{\hb}{\hat{b}}
\newcommand{\hrho}{\hat{\rho}}

\newcommand{\mathat}[1]{\hat{\mathcal{#1}}}

\newcommand{\yuxun}[1]{{{#1}}}
\newcommand{\raw}[1]{{{}}}

\newcommand{\drop}[1]{}

\PassOptionsToPackage{english}{babel}
\begin{document}

\title{Fast Optomechanical Photon Blockade}

\author{Yuxun Ling}
\affiliation{Physics Department,	Blackett Laboratory, Imperial College London, Prince Consort Road, SW7 2BW, United Kingdom}
\author{Sofia Qvarfort}
\affiliation{Nordita, KTH Royal Institute of Technology and Stockholm University, Hannes Alfv\'{e}ns v\"{a}g 12, SE-106 91 Stockholm, Sweden}
\affiliation{Department of Physics, Stockholm University, AlbaNova University Center, SE-106 Stockholm, Sweden}
\author{Florian Mintert}
\affiliation{Physics Department,	Blackett Laboratory, Imperial College London, Prince Consort Road, SW7 2BW, United Kingdom}
\affiliation{Helmholtz-Zentrum Dresden-Rossendorf, Bautzner Landstraße 400, 01328 Dresden, Germany}

\date{\today}

\begin{abstract}
The photon blockade effect is commonly exploited in the development of single-photon sources.
While the photon blockade effect could be used to prepare high-fidelity single-photon states in idealized regimes, practical implementations in optomechanical systems suffer from an interplay of competing
processes.
Here we derive a control scheme that exploits destructive interference of Fock state amplitudes of more than one photon. 
The resulting preparation time for photon-blockaded quantum states is limited only by the optomechanical interaction strength and can thus be orders of magnitude shorter than in existing schemes that achieve photon-blockaded state in the steady state.
\end{abstract}

\maketitle

\section{Introduction}

\yuxun{Single-photon sources are important resources in many fields, including quantum computation~\cite{LO08,WSL+20} and quantum communication~\cite{OFV09}.
Therefore, the photon blockade effect~\cite{ISW+97} has been a research topic of great interest because it can be utilized to realize single-photon sources~\cite{FMV10,WML+16,GKM+17}.
In analogy with the Coulomb blockade of electrons, the photon blockade effect occurs if a single photon in a cavity provides an energetic barrier for a second photon to enter the cavity.}
This phenomenon has been predicted in several systems, such as Kerr-type nonlinear cavity~\cite{HML+18,GL19}, cavity-quantum-electrodynamics systems~\cite{FMV10,BLB+18,HZH+18,TRF+19}, superconducting circuits~\cite{LXM+14,WML+16,GKM+17} and optomechanical systems~\cite{Rabl11,WGL+15,ZLW+18}.

This work specifically considers a generic optomechanical system~\cite{AKM14} consisting of an optical cavity and a mechanical oscillator where the displacement of the mechanical oscillator is linearly coupled to the intensity of the light field inside the cavity.
The system can be experimentally realized in several different ways such as suspended mirror~\cite{VTB+09,GHV+09,KPJ+11}, whispering-gallery-mode microresonators~\cite{DBS+11,VDW+12,DFK+12,JWK+15,MZO+16}, levitated particles~\cite{MMG+08,SLZ+11,MMP+20,SJO+20} and clamped systems~\cite{MHP+11,TDL+11,CAS+11}.

There is substantial literature on the photon blockade effect in optomechanical systems~\cite{Rabl11,SKH+12,LN13,LL13,XLL13,KLM13,KBS+13,LZA+13,WGL+15,ZLW+18}. 
However, existing schemes achieve the photon blockade effect only in the steady state, where pumping of the cavity and photon decay out of the cavity reach equilibrium.
The time scale required for the preparation of a photon-blockaded state is thus at least as long as the lifetime of a photon in the cavity.
In particular, in devices with a high-quality cavity, this poses a severe limitation to the repetition rate at which photon-blockaded quantum states can be prepared.

In order to exploit the photon blockade on shorter time scales, we explore the use of a suitably shaped bichromatic driving of the cavity.
Rather than waiting for the occupation of two-photon states (or states with more photons) to decay, as is typically the case in schemes based on monochromatic driving,
the dynamics resulting from bichromatic driving can lead to destructive interference of different amplitudes of the two-photon state.
We show that optimized bichromatic driving profiles can be used to realize quantum states with a value of the equal-time, second-order autocorrelation function that is three orders of magnitude smaller than what can be achieved with monochromatic driving.

The paper is structured as follows.
In \secname{background}, the model and background are introduced, and in \secname{theoryMethods} the theoretical methods are summarized.
Then, numerical results are shown in \secname{num}.
In particular, the autocorrelation function when a monochromatic driving function is adopted is examined in \secname{num_mono}, where it is shown that the monochromatic driving is insufficient for strong photon blockade at time scales shorter than the dissipation time scale.
In \secname{num_bi}, monochromatic driving is then compared to bichromatic driving with which strong photon blockade is predicted within the same amount of time.
Crucially, in \secname{flatMin} and \secname{highN1}, it is shown that it is possible to increase the duration in which the autocorrelation function stays around its minimum value, as well as increasing the single-photon occupation at the cost of weaker photon blockade.
Finally, limitations in the theory methods are discussed in \secname{discuss}.

\section{Background}\label{sec:background}

\subsection{Optomechanical Photon Blockade}\label{sec:background_1PB}

In this work, we consider an optomechanical system driven by an external coherent light source.
The bare system can be modelled by the Hamiltonian
\be\label{eq:sysHam}
\hat{H}_{OM}&=\omega_c \ha^\dag \ha+\omega_m \hb^\dag \hb-g_0\ha^\dag \ha(\hb^\dag+\hb)\ ,
\ee
with the cavity (mechanical) resonance frequency $\omega_c$ ($\omega_m$),
the annihilation and creation operators $\ha$ ($\hb$) and $\ha^\dag$ ($\hb^\dag$) for the optical cavity field (the mechanical oscillator)
and the single-photon optomechanical coupling strength $g_0$.
The driving Hamiltonian reads
\be\label{eq:drivHam}
\hat{H}_d=\xi^*(t)\ha+\xi(t)\ha^\dag\ ,
\ee
with a complex-valued driving function $\xi(t)$ describing the pumping of the cavity through an external light source.
The temporal shape of the driving function $\xi(t)$ can be chosen with regular pulse-shaping techniques, and the central goal of this paper is the optimization of the function $\xi(t)$ for the realization of single-photon states.

The spectrum of the full system Hamiltonian in Eq.\eqref{eq:sysHam} is anharmonic because of the cubic interaction term (quadratic in optical creation/annihilation operators and linear in mechanical creation/annihilation operators).
This spectrum is most easily understood in the frame defined by the polaron transformation~\cite{Rabl11}
\be
\hat{U}_p=\exp(\frac{g_0}{\omega_m}\ha^\dag \ha(\hb^\dag-\hb))\ .
\ee
In this frame, the undriven system Hamiltonian $\tilde H_{OM}=\hat{U}_p^\dag \hat{H}_{OM}\hat{U}_p$ reads
\be\label{eq:polaronHam}
\tilde H_{OM}=\left(\omega_c-\frac{g_0^2}{\omega_m}\ha^\dag \ha\right) \ha^\dag \ha+\omega_m \hb^\dag \hb\ .
\ee
Instead of the optomechanical interaction term, the Hamiltonian now contains a Kerr-type nonlinearity in the optical subsystem.
Since the eigenstates of the transformed Hamiltonian $\tilde H_{OM}$ are given by the direct product of photonic and phononic Fock states,
the spectrum of the transformed Hamiltonian $\tilde H_{OM}$, and thus of the original Hamiltonian $\hat{H}_{OM}$ is given by
\be\label{eq:unequal_spac}
E_{n,m}=n\brkts{\omega_c-n \frac{g_0^2}{\omega_m}}+m\omega_m\ ,
\ee
where $n$ and $m$ denote the number of photons and phonons respectively.

Crucially for the photon blockade, the optical part of the spectrum is not evenly spaced.
A photon entering an empty cavity is thus resonant with the system if it has the frequency $\omega_c-g_0^2/\omega_m$, while a second photon entering the cavity that is already occupied with one photon would need to have the frequency $\omega_c-3g_0^2/\omega_m$ to be on resonance.
An optomechanical system driven by a light field of frequency $\omega_c-g_0^2/\omega_m$ thus allows a single photon to enter the cavity, but energetically forbids a second photon from entering.

While this qualitative picture provides the correct intuition for photon blockade, the exact mechanism is more involved.
In order to understand the workings in more detail, one also has to appreciate that the polaron transformation not only makes the undriven system non-interacting but also modifies the nature of the driving.
Although the driving in the laboratory frame (Eq.~\eqref{eq:drivHam}) results in the excitation and de-excitation of only photons,
the driving Hamiltonian in the polaron frame
\be
\hat{U}_p^\dag \hat{H}_d(t)\hat{U}_p=\xi^*(t)\ha \exp(\frac{g_0}{\omega_m}(\hb^\dag-\hb))+h.c.\ ,
\label{eq:drivepolaron}
\ee
is expressed in terms of photon creation and annihilation operators dressed with a displacement of the mechanical degree of freedom.
The drive can thus not only (de)excite the quantum mechanical light field in the cavity, but it can also create or annihilate phonons.
The energetic violation in creating a second photon that is at the core of the photon blockade can thus be jeopardized by the possibility of depositing excess energy into the mechanical degree of freedom.

The excitation or de-excitation of individual phonons during the excitation of a cavity photon is only negligible if the corresponding energetic violation is large as compared to the driving amplitude, i.e. if $\abs{\omega_m-2g_0^2/\omega_m}\gg \abs{\xi(t)}$.

So far, the discussion has assumed an idealized cavity with a perfectly sharp resonance frequency. However, since any cavity has a finite linewidth $\kappa$, which can affect the suppression of otherwise energetically forbidden processes, the additional condition $\abs{\omega_m-2g_0^2/\omega_m}\gg \kappa$ is required for the photon blockade effect.

The driving Hamiltonian in the polaron frame (Eq.~\eqref{eq:drivepolaron}) includes processes with the creation or annihilation of any number of phonons, and any such process can be energetically suppressed if conditions like the ones formulated above for the creation or annihilation of a single phonon are satisfied.
In particular, to achieve a strong interaction between the cavity field and the mechanical oscillator, an unrealistically large number of such constraints needs to be satisfied.
It is much more practical to require a weak interaction, i.e. $g_0\ll\omega_m$, so that processes including the creation or annihilation of more than a single phonon in Eq.~\eqref{eq:drivepolaron} are negligible.

The realization of photon blockade thus requires sufficiently strong coupling $g_0$ in order to realize an anharmonic spectrum, but at the same time sufficiently weak coupling in order to suppress the creation or annihilation of more than one phonon.

Since the experimental realization of a system with a clear separation of the relevant scales -- 
(i) the mechanical resonance frequency $\omega_m$,
(ii) the Kerr-type nonlinearity $g_0^2/\omega_m$, and
(iii) the cavity linewidth $\kappa$ or the driving amplitude $|\xi(t)|$ --
is extremely challenging,
any practical driving scheme needs to work also outside such an idealized regime.
Existing schemes relying on monochromatic driving can realize a sizeable occupation of the single-photon Fock state and simultaneous suppression of the two-photon Fock state only in the steady state when the effect of external driving and system losses have come to an equilibrium.
The necessity to wait for such an equilibration results in a slow process that limits the efficiency of state preparation.

The goal of this paper is the design of bichromatic driving schemes that help to realize photon-blockaded states of the light field in short time scales.
The use of bichromatic driving allows us to exploit the interference of different absorption processes.
A suitably chosen driving function $\xi(t)$ can thus lead to destructive interference of different paths towards the population of the two-photon Fock state.
This maximizes the strength of photon blockade in an optomechanical system.

\subsection{Photon Blockade and Autocorrelation}

It is helpful to identify a quantifier for the strength of photon blockade. Such a quantifier is the equal-time second-order autocorrelation function, which is given by
\be\label{eq:g20_exact}
g^{(2)}(t)=\frac{\langle\hat n^2\rangle-\langle\hat n\rangle}{\langle\hat n\rangle^2}\ ,
\ee
with the photon-number operator $\hat n=\ha^\dag\ha$ and the expectation operator $\avg{\cdot}$ with respect to the system state $\hrho(t)$.
Intuitively, the autocorrelation function $g^{(2)}(t)$ describes how likely it is to find more than one photon in the system at the time $t$.
Therefore, the stronger the photon blockade, the lower the value of the autocorrelation function $g^{(2)}(t)$ is, and for full photon blockade, the autocorrelation vanishes.

Despite its simple form, the time evolution of the autocorrelation function is not easy to analytically determine for optomechanical systems.
This is because the system dynamics cannot be solved exactly due to the cubic interaction term.
To address the problem, this work considers the photon blockade effect in a perturbative regime in which the occupation probability of Fock states containing three or more photons is negligible as compared to the single-photon and two-photon occupations.
When such higher-photon occupations are suppressed, the autocorrelation function in \eq{g20_exact} can be expressed as
\be\label{eq:g20_smallHS}
g^{(2)}(t)\approx\frac{2p_2(t)}{(p_1(t)+2p_2(t))^2}:=\tilde g^{(2)}(t)\ ,
\ee
in terms of the occupation probabilities $p_1(t)$ and $p_2(t)$ of the single- and two-photon Fock state respectively.
In the remainder, it is shown that the approximate autocorrelation function $\tilde g^{(2)}(t)$ can be analytically determined for optomechanical systems, and the photon blockade effect are then studied based on the results.
As we show in the upcoming sections, optimal bichromatic driving schemes can help to prepare quantum states with a value of the exact $g^{(2)}$-function as low as $10^{-4}$ and a time scale as short as five periods of mechanical motion.

\subsection{Dissipative Dynamics in Optomechanical Systems}\label{sec:dynamicsDfn}

All quantum systems are coupled to their environment, which gives rise to dissipative processes.
There has been impressive progress in the reduction of mechanical dissipation in many optomechanical experiments~\cite{CAS+11,VDW+12,MRL+20,RMM+20},
but the necessity to open the cavity to inject light implies a finite rate of photon loss from the cavity.
The following discussion thus takes into account optical loss but neglects mechanical dissipation.
The effects of mechanical dissipation are explored in Sec.~\ref{sec:discuss}.

Optical loss for any state $\hrho$ at rate $\kappa$ can be modelled with the Lindbladian $\mathat{L}_c$ satisfying the relation
\be
\mathat{L}_c\hrho=\frac{\kappa}{2}\left(2\hat{a}\hrho \hat{a}^\dag-\left(\hat{a}^\dag \hat{a}\hrho+\hrho \hat{a}^\dag \hat{a}\right)\right)\ .
\ee
The complete Master equation $\dot \hrho=\mathat{L}\hrho$ of the system (in the original frame instead of the polaron frame) including drive and dissipation is then given by the generator $\hat{\cal L}$ of the system dynamics that satisfies the equation
\be\label{eq:MasterEq}
\mathat{L}\hrho=-i[\hat{H}_{OM}+\hat{H}_d,\hrho]+\mathat{L}_c(\hrho)\ ,
\ee
with the system Hamiltonian $\hat{H}_{OM}$ as defined in \eq{sysHam} and the driving Hamiltonian $\hat{H}_{d}$ as defined in \eq{drivHam}.
This can be taken as a starting point to solve for the time dependence of the photon statistics in optomechanical systems.

\section{Solving the system dynamics}\label{sec:theoryMethods}

\yuxun{While steady-state properties can be inferred without solving the system dynamics, the goal of this work requires an explicit solution of the system's equations of motion.
The dynamics of the undriven and dissipation-less system can be solved exactly~\cite{MMT97,BJK97}.
However, even though the dynamics of some observables can be constructed exactly also in the presence of drive~\cite{LM18,LM21} and dissipation~\cite{QVB+21},
this is not the case for the photon occupations of interest here.
It is thus necessary to construct these quantities perturbatively.
\raw{Perturbative solutions are generally limited to short-time dynamics, but since the present goal is to identify how to realize photon-blockaded states as quickly as possible, this is hardly a constraint.}
Perturbative solutions can then enable the identification of realizing photon-blockaded states as quickly as possible, which is the goal of this work.}

To solve the dynamics, it is convenient to note that the generator $\mathat{L}$ in \eq{MasterEq} can be separated into a photon-number-conserving term $\mathat{L}_0$ with
\be
\mathat{L}_0\hrho&=-i[\hat{H}_{OM},\hrho]-\frac{\kappa}{2}(\ha^\dag \ha\hrho+\hrho \ha^\dag \ha)\ ,
\ee
and a term $\mathat{L}_1$ satisfying the equation
\be\label{eq:MasterPhotVar}
\mathat{L}_1\hrho&=-i[\hat{H}_d,\hrho]+\kappa\ha\hrho \ha^\dag\ ,
\ee
which captures processes in which the photon number is not conserved.
In the interaction picture, the time-evolution of the system can then be described by the following equation
\be
\hrho(t)=\mathat{S}_0(t)\mathat{S}_1(t)\hrho_0\ ,
\ee
with the propagator $\mathat{S}_0$ describing the dynamics induced by the photon-conserving term $\mathat{L}_0$, the propagator $\mathat{S}_1$ describing the photon-varying processes induced by the photon-varying term $\mathat{L}_1$ in the frame defined by $\mathat{S}_0$, and $\hrho_0$ being the initial state of the system.

The dynamics induced by the photon-conserving term $\mathat{L}_0$ can be constructed exactly by solving the differential equation 
\be
\frac{d}{dt}\mathat{S}_0(t)\hrho_0= \mathat{L}_0\mathat{S}_0(t)\hrho_0\ ,
\ee
for any initial state $\hrho_0$ of the system. 
The full analytic expression of the propagator $\mathat{S}_0$ is lengthy but interested readers can find it together with its derivation in \apdxname{sysDymSoln}.

The remainder of the dynamics must be solved perturbatively.
The generator $\tilde{\mathcal M}$ of the rest of the system dynamics in the frame defined by $\mathat{S}_0$ reads
\be\label{eq:SvDym}
\tilde{\mathcal M}(t)=\mathat{S}_0^{-1}(t)\mathat{L}_1(t)\mathat{S}_0(t)\ .
\ee
The propagator $\mathat{S}_1$ induced by the generator $\tilde{\mathcal M}$ can be constructed perturbatively in the driving strength $\abs{\xi(t)}$ and the optical decay $\kappa$ using the Neumann series~\cite{Neu77} which read as $\mathat{S}_1=\sum_{k=0}^\infty\mathat{S}^{(k)}_1$.
The zeroth order term $\mathat{S}^{(0)}_1(t)=\hat{\mathbbm{1}}$ is the identity map, and the $k$-th order term $\mathat{S}^{(k)}_1(t)$ (for $k>0$) reads
\be\label{eq:Neumann}
\mathat{S}^{(k)}_1(t)=\int_{\bm{t^{(k)}}\in\mathbb{T}}\brkts{\prod_{j=1}^k \tilde{\mathcal M}(t_j)}d\bm{t^{(k)}}\ ,
\ee
where $\bm{t^{(k)}}=(t_1,\hdots,t_k)$ is the vector of time-ordered time variables $t_1>\hdots>t_k$ and $\mathbb{T}$ is the hyperpyramidal domain $t>t_1>\hdots>t_k>0$.

Due to the time dependence of the propagator $\mathat{S}_0$, it is generally not possible to evaluate the integrals in Eq.~\eqref{eq:Neumann} analytically.
However, in order to construct the time-dependence of the probabilities $p_1$ and $p_2$ with corrections up to their leading order in $\abs{\xi}/\omega_m$, only the terms $\mathat{S}^{(k)}_1(t)$ with $k\le 4$ are required.
Furthermore, all terms in the $\mathat{S}^{(k)}_1(t)$ that are at least quartic in $g_0/\omega_m$ can be neglected.
In these approximations, the probabilities $p_1$ and $p_2$ can indeed be constructed analytically for any driving function $\xi(t)$ that is given as a Fourier series.
This is discussed in more detail in~\apdxname{finRes_theory}.

\section{Results}\label{sec:num}

We strive in the following for the identification of driving patterns such that the system evolves towards a state with a minimal value of the autocorrelation $g^{(2)}$ at a desired point in time.
This is achieved using the analytic dependence of the approximate autocorrelation function $\tilde g^{(2)}$ on the driving profile $\xi(t)$ derived in Sec.~\ref{sec:theoryMethods}, 

Since most of the earlier works focused on monochromatic driving
\be\label{eq:monoDriv}
\xi_1(t)=\epsilon e^{-i(\omega_c+\Delta)t}\ ,
\ee
with driving strength $\epsilon$ and detuning $\Delta$ from the cavity resonance frequency,
this situation is relevant for comparison.
The main focus of the subsequent discussion, however, is on bichromatic driving
\be\label{eq:biDriv}
\xi_2(t)=\epsilon_1 e^{-i(\omega_c+\Delta_1)t}+\epsilon_2 e^{i\psi}e^{-i(\omega_c+\Delta_2)t}\ ,
\ee
with two driving amplitudes $\epsilon_j$ and corresponding detunings $\Delta_j$ as well as a phase difference $\psi$ between the two driving components.

The construction of the optimized driving patterns is based on the system ground state $\hrho_0=\ketbra{0}\otimes\ketbra{0}$ as the initial state.
Even though the approximate autocorrelation function $\tilde g^{(2)}$ is given as an analytic function of all system parameters including the Fourier components of the driving profile $\xi$, the actual optimization of the drive cannot be performed analytically.
In all the subsequent examples, numerical optimization is performed with Mathematica~\cite{Mathematica}.

The validity of all results obtained based on perturbative treatments is verified through comparison with numerically exact solutions.
Numerical simulations are performed with Qutip~\cite{Qutip1, Qutip2} in a Hilbert space truncated to include up to $6$ photons and $15$ phonons.

Since the explicit shape of the driving patterns identified as optimal is not necessarily insightful on its own, it is not specified directly in the subsequent discussion.
For completeness, however, the optimized numerical values of the parameters in the driving functions $\xi_1$ and $\xi_2$ can be found in \apdxname{params}.

\subsection{Photon Blockade for Monochromatic Driving}\label{sec:num_mono}

\begin{figure}[t]
   \centering
   \includegraphics[keepaspectratio, width=0.45\textwidth]{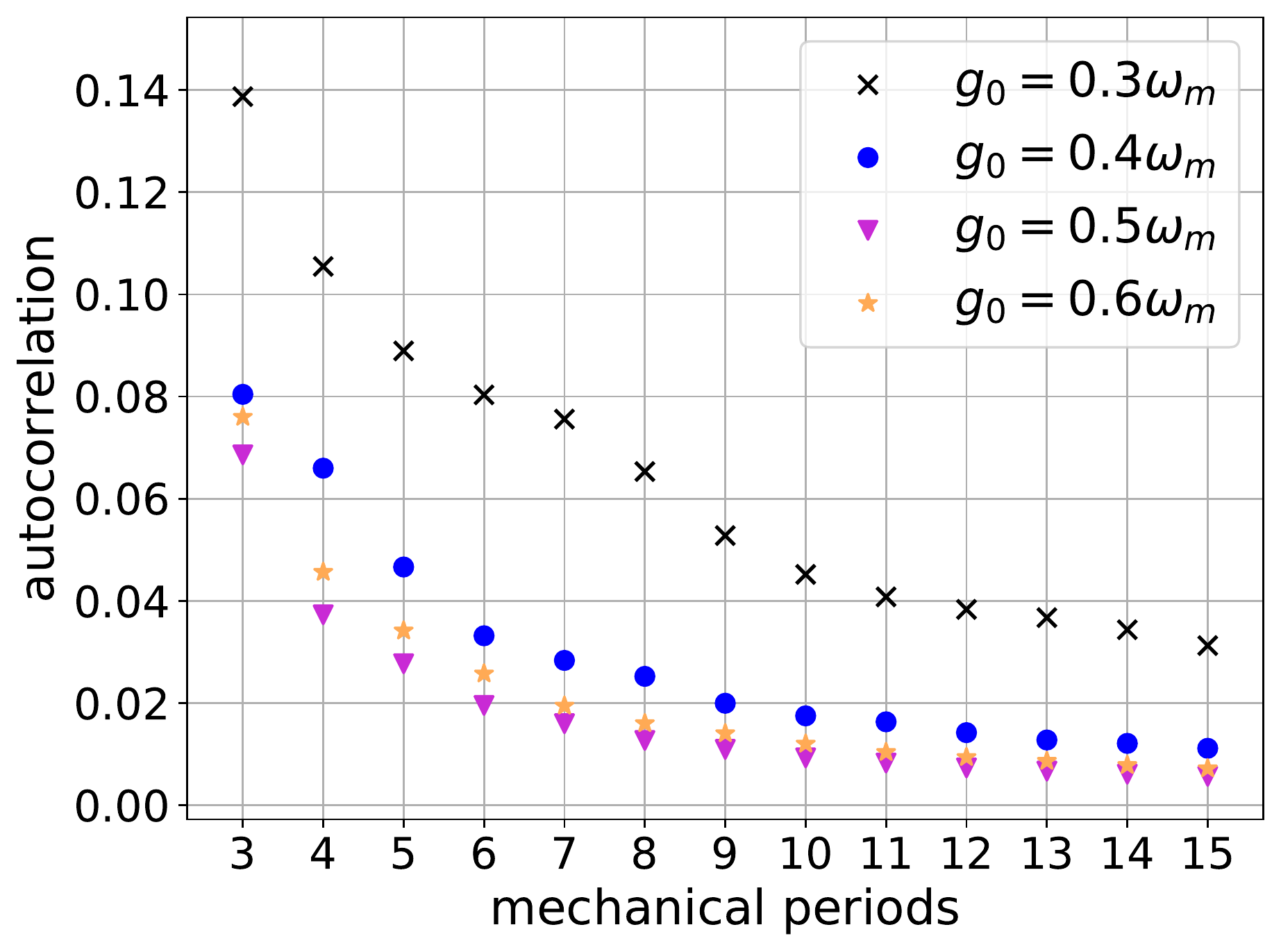}
   \caption{The value of the autocorrelation function $g^{(2)}$ as a function of the target time $t_{op}$ in terms of mechanical periods ($T=2\pi/\omega_m$) when the approximate autocorrelation function $\tilde{g}^{(2)}(t_{op})$ is minimized with respect to the driving detuning $\Delta$ at the considered time $t_{op}$.
Different values of the optomechanical coupling strength $g_0=0.3\omega_m$ (cross), $g_0=0.4\omega_m$ (circle), and $g_0=0.5\omega_m$ (triangle) and $g_0=0.6\omega_m$ (star) are considered.
The driving strength is $\epsilon=\omega_m/200$ and cavity decay rate is $\kappa=0.02\omega_m$.
The minimized autocorrelation function decreases when either the driving strength or the evolution time increases, but strong photon blockade ($g^{(2)}\sim10^{-4}$) cannot be observed within the range of parameters considered.
}
   \label{fig:mono_g2_vs_t} 
 \end{figure}

For comparison with the results obtained with bichromatic driving discussed below in \secname{num_bi}, this section is focused on the photon-blockaded states that can be realized with monochromatic driving.
In particular, it is can be found that, within the perturbative regime ($(g_0/\omega_m)^4,(\epsilon/\omega_m)^2\ll1$), monochromatic driving in \eq{monoDriv} is insufficient for strong photon blockade in the time limit $t\ll1/\kappa$.
\Figname{mono_g2_vs_t} plots the values of the autocorrelation function $g^{(2)}(t_{op})$ from the full numerical simulation as a function of the target time $t_{op}$.
For each data point, the driving detuning $\Delta$ is optimized for minimal approximate autocorrelation function $\tilde{g}^{(2)}(t_{op})$ at the target time $t_{op}$.
In addition, a range of coupling strength ($g_0=0.3\omega_m,\hdots,0.6\omega_m$) is also considered while other parameters are kept constant $(\epsilon,\kappa)=(\omega_m/200,\omega_m/50)$. 

The strength of photon blockade depends on the optomechanical coupling strength $g_0$.
This can be confirmed by the data points corresponding to different coupling strength $g_0$ for each fixed target time $t_{op}$ in \figname{mono_g2_vs_t}. 
The minimized values of the autocorrelation function decrease as the coupling strength $g_0$ increases until it reaches the value $g_0=0.5\omega_m$.
The main reason for the increase in the autocorrelation is that the energy difference between different transitions in the unequally spaced energy spectrum (\eq{unequal_spac}) is proportional to the square of the coupling strength (i.e. $E_{n+1,m}-E_{n,m}\propto g_0^2$).
As the coupling strength increases further (e.g. $g_0\ge0.6\omega_m$, starred points in \figname{mono_g2_vs_t}), the energy required to pump a single photon into the cavity is eventually approached by the energy required to pump a second photon and an extra phonon into the cavity (i.e. $E_{2,1}-E_{1,0}=E_{1,0}-E_{0,0}$), and thus the photon blockade effect vanishes~\cite{Rabl11}.
This can be confirmed by the increase in the autocorrelation function for larger coupling strength ($g_0>0.5\omega_m$).
The increase in the autocorrelation for larger coupling strength is also consistent with the fact that the assumption of small coupling strength ($(g_0/\omega_m)^4\ll1$) is violated close to the value $g_0=0.5\omega_m$, but the errors only contribution to a small portion of the increase.\footnote{The error of the theoretical values relative to the numerical values increases from $1.4\%$ to $5.1\%$ as the target time $t_{op}$ increases from $3T$ to $15T$.
However, the relative difference between the autocorrelation for coupling strength with values $g_0=0.5$ and $g_0=0.6$ ranges from $11\%$ to $32\%$ which are one order of magnitude larger than the relative difference between analytical results and numerical results.}

Furthermore, \figname{mono_g2_vs_t} also shows that monochromatic driving is insufficient for strong photon blockade for any target time within the first $15$ mechanical periods.
By looking at the data points corresponding to different target time $t_{op}$ for each fixed coupling strength $g_0$, it can be observed that the minimized values of the autocorrelation function $g^{(2)}(t_{op})$ decrease monotonically as the target time $t_{op}$ increases.
However, the best value ($\sim10^{-2}$) of the autocorrelation function that can be obtained within the first $15$ mechanical periods ($T=2\pi/\omega_m$) is still far from the value ($g^{(2)} \lesssim 10^{-4}$) that can be achieved using a bichromatic drive.

\subsection{Fast Photon Blockade for Bichromatic Driving}\label{sec:num_bi}

\begin{figure}[t]
   \centering
   \includegraphics[keepaspectratio, width=0.45\textwidth]{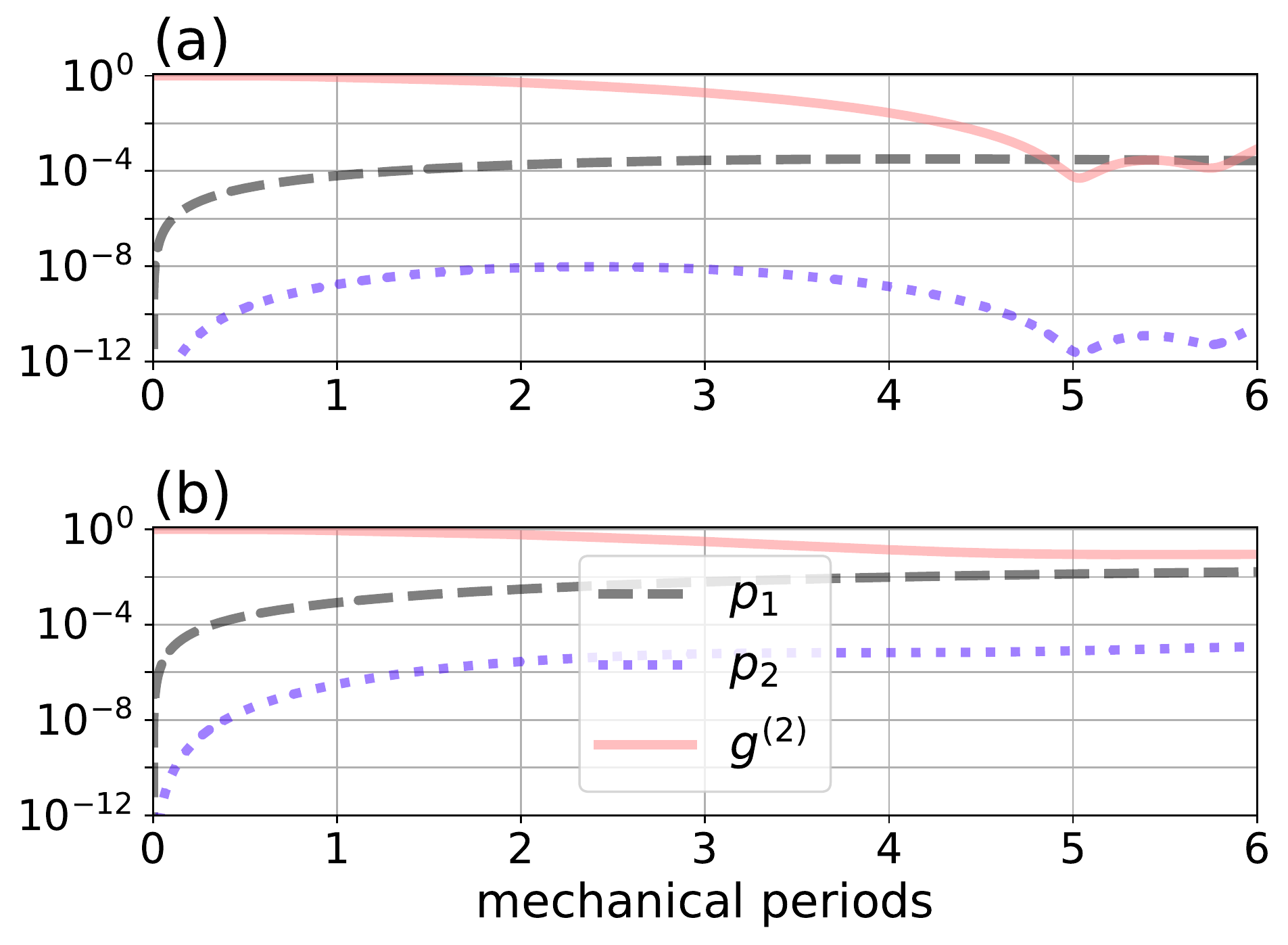}
   \caption{Single-photon occupation $p_1$ (dashed), two-photon occupation $p_2$ (dotted) and autocorrelation function $g^{(2)}$ (solid) as functions of time for parameters that minimize the autocorrelation function at $5$-th mechanical period when the parameters have the values $\epsilon=\omega_m/200$, $g_0=0.3\omega_m$, $\kappa=0.02\omega_m$, and the driving detuning and phases are optimized. Inset (a) corresponds to the dynamics when the driving is bichromatic and strong photon blockade ($g^{(2)}\sim10^{-4}$) can be observed at $5T$. Inset (b) corresponds to the dynamics when the driving is monochromatic and the lowest value of the autocorrelation function only reaches $10^{-1}$. }
   \label{fig:bichrom_raw} 
 \end{figure}

It is possible to induce a stronger fast photon blockade given an optimized bichromatic drive.
To compare a bichromatic drive with a monochromatic drive, the time-evolution of the autocorrelation function $g^{(2)}$ is plotted in \figname{bichrom_raw} as a function of time $t$. 
\Figname{bichrom_raw}(a) shows the autocorrelation $g^{(2)}$ for the bichromatic drive, while \figname{bichrom_raw}(b) shows the autocorrelation $g^{(2)}$ for a monochromatic drive. 
In addition to the autocorrelation $g^{(2)}$, the single- and two-photon occupation numbers $p_1$ and $p_2$ are plotted to demonstrate the difference in the time evolution of the optical subsystem between the two cases. The cavity decay rate  is set to $\kappa=\omega_m/50$, as in \secname{num_mono}.
The coupling strength is fixed to $g_0=0.3\omega_m$, which is less than the optimal value ($g_0\approx0.5\omega_m$) in \figname{mono_g2_vs_t}.\footnote{This is to ensure that the assumption of small coupling strength ($(g_0/\omega_m)^4\ll1$) is satisfied with an error less than $0.01$.} The driving strength for each component is kept constant as $\epsilon_1=\epsilon_2=\omega_m/200$. In both plots, the detunings take the optimized values $\Delta=-0.040\omega_m$, $\Delta_1=-0.019\omega_m$ and $\Delta_2=-0.028\omega_m$ and the relative phase of the bichromatic drive takes the optimized value $\psi=2.8$ so that the approximate autocorrelation $\tilde{g}^{(2)}$ reaches its minimum around $t_{op} = 5T$. This target time was chosen because it is close to the minimal time required to achieve the autocorrelation value $g^{(2)}(t_{op})\lesssim10^{-4}$. The optimized values can be found in \apdxname{params}. 

From \figname{bichrom_raw}, it can be seen that much lower values of the autocorrelation $g^{(2)}$ can be achieved with the bichromatic drive. The evolution plotted in \figname{bichrom_raw}(a) shows a second-order autocorrelation of value $g^{(2)}(5T)\approx5.8\times10^{-5}$ at the end of the fifth mechanical period, which is three orders of magnitude smaller than the best possible value for a monochromatic driving (see \secname{num_mono}) $g^{(2)}(5T)\approx8.9\times10^{-2}$ shown in \figname{bichrom_raw}(b).
The significant enhancement in the strength of the photon blockade effect demonstrates the benefit of adopting an optimized bichromatic driving. 
Furthermore, it can be observed that the single-photon occupation $p_1$ also increases monotonically in both subplots but the two-photon occupation only shows a rapid decrease near the $5$-th mechanical period when a bichromatic driving is adopted.
This supports the argument in \secname{background} that the entry of the second drive induces strong destructive interference only between different transition paths to the two-photon Fock state without affecting the single-photon occupation. It is also worth noticing that the low autocorrelation value $g^{(2)}(5T)\approx5.8\times10^{-5}$ is achieved given a coupling strength $g_0$ weaker than the optimal value indicated by \figname{mono_g2_vs_t}, which suggests a further advantage of adopting a bichromatic driving.

It can also be noticed that in \figname{bichrom_raw}, the single-photon population $p_1$ for the two subplots are different.
However, the comparison of the intracavity photon numbers  is not insightful.
This is because the compound driving strength $\abs{\xi_2(t)}$ (see \eq{biDriv}) of a bichromatic drive is dynamic in contrast to the time-independent driving strength $\abs{\xi_1(t)}=\epsilon$ of a monochromatic drive.
Therefore, the compound driving strength $\abs{\xi_2(t)}$ of a bichromatic drive can be much weaker than that of a monochromatic drive even if the driving strengths of the components are equal (i.e. $\epsilon_1=\epsilon_2=\epsilon$).
On the other hand, the autocorrelation $g^{(2)}$ does not suffer from the same problem.
As long as the driving strengths for the two driving components are equal and weak (i.e. $\epsilon_1=\epsilon_2\ll \omega_m$), the value of the approximate autocorrelation function $\tilde{g}^{(2)}$ to leading order becomes independent of the driving strength.
Therefore, the value of the exact autocorrelation function $g^{(2)}$ is independent of the driving strength given the validity of the perturbative method. Further discussion of this point can be found in \secname{modelInaccu}.
\begin{figure}[t]
   \centering
   \includegraphics[keepaspectratio, width=0.45\textwidth]{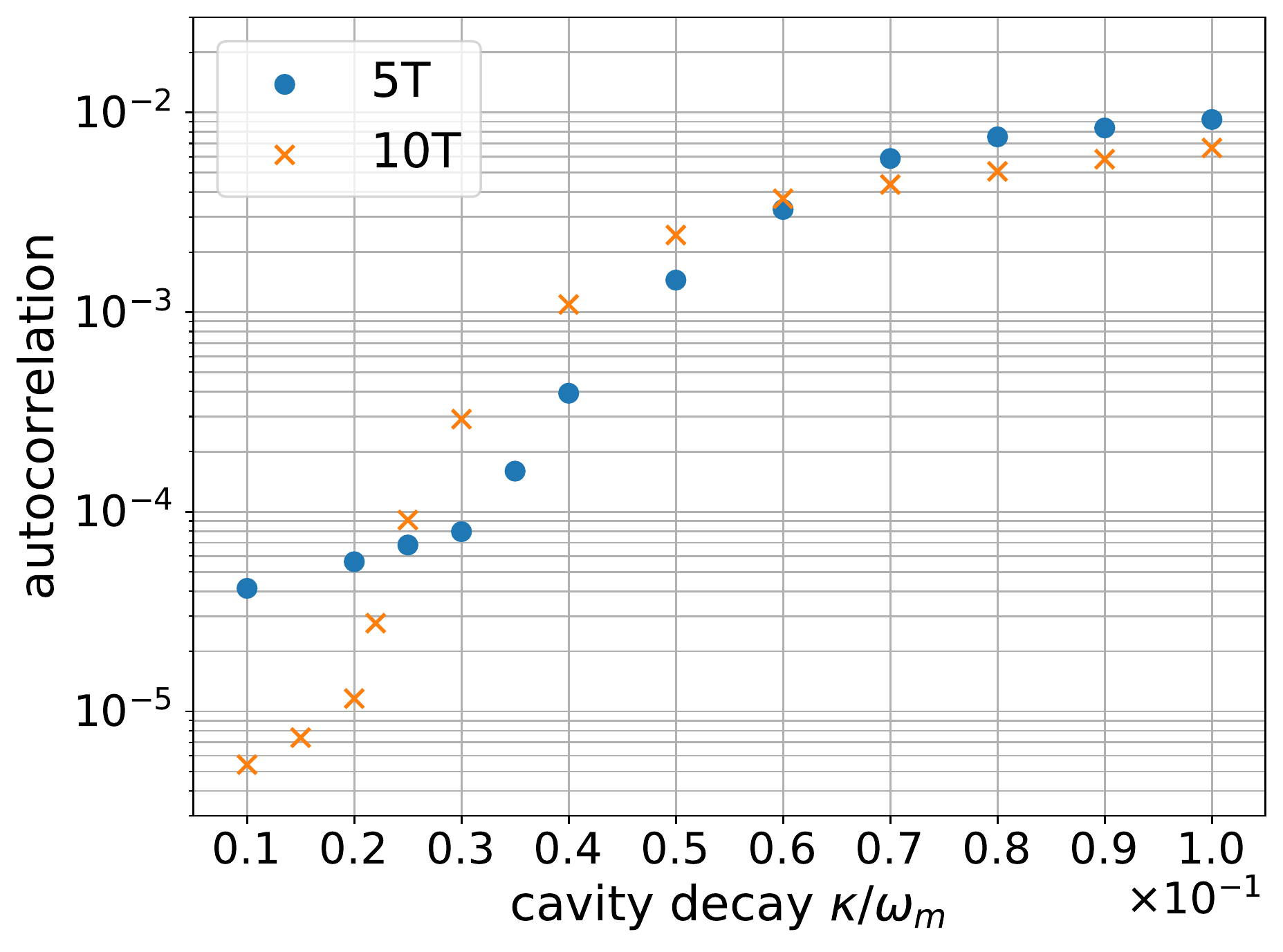}
   \caption{The minimal value of autocorrelation function against the relative cavity decay rate $\kappa/\omega_m$ when $t=5T$ (circle) and $t=10T$ (cross). 
The driving strength is $\epsilon=\omega_m/200$ and optomechanical coupling rate is $g_0=0.3\omega_m$
The minimal values of the autocorrelation function increase as the cavity decay increases, but do not have a monotonic dependence on the evolution time.}
   \label{fig:bi_g2_vs_kap} 
 \end{figure}

Although the driving function is the only degree of freedom to be optimized, other system parameters also need to be kept within a suitable range for the strong photon blockade to occur. Especially, the cavity decay rate has a strong influence on $g^{(2)}$. 
To explore the behaviour of the $g^{(2)}$-function as the cavity decay $\kappa$ increases, the minimized values of the autocorrelation function $g^{(2)}(t_{op})$ are plotted in  \figname{bi_g2_vs_kap}. Two different evolution time are considered, namely $t_{op}=5T$ (blue circles) and $t_{op}=10T$ (yellow crosses). The optomechanical coupling is kept at $g_0 = 0.3\omega_m$ and the driving is $\epsilon_1 = \epsilon_2 = \omega_m/200$ as before. For each point, $\Delta_1$, $\Delta_2$ and $\psi$ have been optimized such that $g^{(2)}$ reaches its lowest value. 

From studying \figname{bi_g2_vs_kap}, it is clear that the minimal value of the autocorrelation function $g^{(2)}$ stays below $10^{-4}$ as long as the cavity linewidth remains below $\kappa \sim 0.03\omega_m$. 
However, as $\kappa$ becomes comparable to the optomechanical coupling $g_0$, the unequal spacings between different energy levels, as shown in \eq{unequal_spac}, are no longer distinguishable by the cavity, and thus the photon blockade effect eventually vanishes for large $\kappa$. 

It is also noticeable from \figname{bi_g2_vs_kap} that a longer evolution time does not necessarily lead to a lower minimal $g^{(2)}$. This is in sharp contrast with the monochromatic case, as seen in \figname{bichrom_raw}. 
Such a feature can be explained by the fact that the occurrence of the minimum of $g^{(2)}$ depends on the destructive interference between the two driving fields and it is not guaranteed that a longer evolution time enhances the destructive interference between the two driving tones.

While it is generally important to minimize the autocorrelation function, there are other aspects that can be further improved. They are explored in the following sections.

\subsection{Extending the Lifetime of Photon Blockade}\label{sec:flatMin}

The results presented in the previous section show the desired low values of the autocorrelation function $g^{(2)}$, which indicate strong photon blockade.
Besides strong photon blockade, it is also helpful to extend the duration for which the photon-blockaded state can be accessed or equivalently, the duration in which the autocorrelation function stays at its minimum value.
In this section, it is shown that the maximization of this duration can be achieved at the same time when the strength of photon blockade is maximized.

A broader minimum of the autocorrelation can be achieved by minimizing the following target function
\be\label{eq:targetFunc1}
\eta_1=\biggl(1+w_d\frac{d}{dt}+w_s\frac{d^2}{dt^2}\biggr)\tilde{g}^{(2)}(t_{op})\ ,
\ee
with scalar weights $w_d$ and $w_s$.
The minimization of the first-order derivative then guarantees that the local minimum of the autocorrelation function is located in the vicinity of the given time $t_{op}$.
A minimal second-order derivative then assures that the vanishing first-order derivative stays at a low value around the given time $t_{op}$.

\begin{figure}[t]
   \centering
   \includegraphics[keepaspectratio, width=0.45\textwidth]{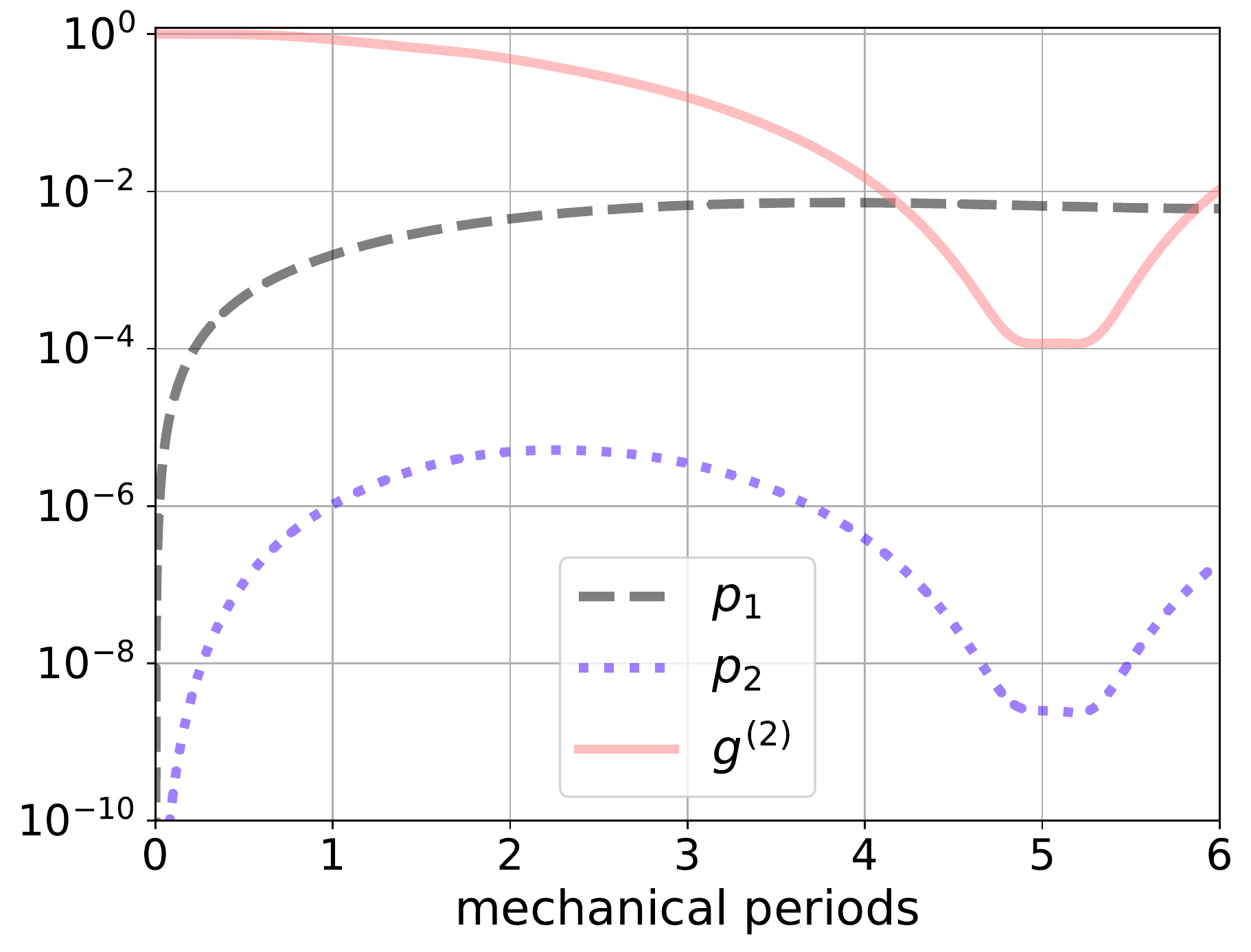}
   \caption{Single-photon occupation $p_1$ (dashed), two-photon occupation $p_2$ (dotted) and autocorrelation function $g^{(2)}$ (solid) as functions of time for parameters that minimize the target function \eq{targetFunc1} at $5$-th mechanical period when the parameters have the values $\epsilon=\omega_m/200$, $g_0=0.3\omega_m$ and $\kappa=0.02\omega_m$. 
The autocorrelation function stays around its minimum for approximately $0.2$ mechanical periods.}
   \label{fig:bi_g2_vs_t_flat} 
 \end{figure}

\Figname{bi_g2_vs_t_flat} demonstrates that this choice of target function produces the desired result.
A bichromatic driving is optimized for a flat minimum with fixed weights $w_d=1$ and $w_s=10$ and the other system parameters are chosen to be $(\epsilon_1,\epsilon_2,g_0,\kappa,t_{op})=(\omega_m/200,\omega_m/200,3\omega_m/10,\omega_m/50,5T)$ in accordance with \figname{bichrom_raw}.
The weight $w_s$ for the second-order derivative is chosen to be larger in order to prioritize the maximization of the width of the minimum.
As can be noted from \figname{bi_g2_vs_t_flat}, the optimized driving in \figname{bi_g2_vs_t_flat} makes the autocorrelation stay within $\pm5\%$ of its value at the target time $t_{op}=5T$ for $0.38$ mechanical periods as compared to $1.1\times10^{-2}$ mechanical periods in \figname{bichrom_raw}.
The improvement is more than one order of magnitude, which suggests that the target function $\eta_1$ in \eq{targetFunc1} correctly predicts the desired driving function that widens the minimum.

There are however drawbacks to the optimization for broader minima.
As a trade-off, the value of the autocorrelation function $g^{(2)}(5T)=1.2\times10^{-4}$ at is slightly higher as compared to the result in \figname{bichrom_raw} ($g^{(2)}(5T)\approx5.8\times10^{-5}$).
However, this still suggests strong photon blockade (i.e. $g^{(2)}\sim10^{-4}$).
The trade-off also indicates that the increase in the length of time in which the autocorrelation function stays around its minimum is not unlimited, but rather depends on the threshold value of the autocorrelation function for strong photon blockade.

\subsection{Optimization for Higher Single-Photon Occupation}\label{sec:highN1}

While it is important to maximize the strength of the photon blockade effect by minimizing the $\tilde{g}^{(2)}$ function, maximizing the single-photon population $p_1$ at the same time is also important if a single-photon source is targeted.
This can be done by minimizing the following target function
\be\label{eq:targetFunc2}
\eta_2=\biggl(1+w_d\frac{d}{dt}+w_s\frac{d^2}{dt^2}\biggr)\tilde{g}^{(2)}(t_{op})+\frac{w_1}{p_1(t_{op})}\ ,
\ee
with scalar weight $w_1$.
The target function $\eta_2$ in \eq{targetFunc2} is the target function $\eta_1$ in \eq{targetFunc1} plus a second term that maximizes the single-photon population.
Although the multi-photon populations are automatically minimized whenever the single-photon population approaches unity, this is not the case in the perturbative regime of weak driving.
Therefore, the autocorrelation function still requires minimization at the same time when the single-photon population is maximized.
The target function $\eta_2$ also assumes that the single-photon population only varies slowly over the interval in which the autocorrelation function stays near its minimum.
This assumption is justified by the weak driving strength.

For better comparison with the results presented in \secname{flatMin}, it is helpful to assume the system parameters $(\kappa,g_0,t_{op})=(\omega_m/50,3\omega_m/10,5T)$ and the weights $(w_d,w_s)=(1,10)$.
The weight $w_1$ for the term that is inversely proportional to the single-photon population $p_1$ is taken to be the value $w_1=\epsilon_1^2/10$ which is proportional to the square strength so that the term becomes unitless to leading order.
By minimizing the target function $\eta_2$ with respect to the driving strengths ($\epsilon_1,\epsilon_2$), the detunings ($\Delta_1, \Delta_2$) and the phase $\psi$, the single-photon occupation at fifth mechanical period increases to $p_1(5T)\approx9.4\times10^{-3}$ as compared to the value $p_1(5T)\approx6.6\times10^{-3}$ in \figname{bi_g2_vs_t_flat}.
As a trade-off, the value of the autocorrelation function $g^{(2)}(5T)$ also increases from $1.2\times10^{-4}$ to $1.6\times10^{-4}$ but both values are of the same order of magnitude which suggests strong photon blockade.
Similarly, the length of time at which the autocorrelation function stays within $\pm5\%$ of the latter value only slightly reduces to $0.37$ mechanical periods from $0.38$ mechanical periods.
Within this interval, the change in the single-photon population is also less than $5\%$ of its value at the fifth mechanical period, which supports the slow-varying assumption discussed earlier in the section.

The photon population in the cavity is ultimately bounded by the maximal pumping power, and thus the single-photon population can be further enhanced by increasing the driving strength in addition to minimizing the target function \eq{targetFunc2}.
For example, with the optimized driving parameters, the single-photon occupation can be enhanced to exceed the vacuum occupation (i.e. $p_1>0.5$) if the pumping power is increased by $10$-times.
However, such modification can lead to a decrease in the accuracy of the perturbative model as is discussed in \secname{modelInaccu}.

\section{Discussion}\label{sec:discuss}

In this section, we discuss the\raw{ effect of several imperfections in the theoretical model that was used in earlier sections.
For example, the interaction between the mechanical oscillator and the environment can be non-negligible and the perturbative assumptions in the theory fail when the system is driven too strongly.
These inaccuracies in the model are numerically quantified by the difference between the values of the correlation function when the corresponding imperfection is present and when it is omitted.}
\yuxun{assumptions made when deriving the main result. 
For example, the interaction between the mechanical oscillator and the environment can be non-negligible and the perturbative assumptions in the theory fail when the system is driven too strongly.
These differences are quantified by numerically computing the autocorrelation function and comparing it with the analytic approximate function $\tilde{g}(t)$ shown in Eq. \eqref{eq:g20_smallHS}. }
For consistency and unless otherwise specified, this section adopts the system parameters $(\epsilon_1,\epsilon_2,g_0,\kappa,t_{op})=(\omega_m/200,\omega_m/200,3\omega_m/10,\omega_m/50,5T)$ and the driving parameters as used in \figname{bi_g2_vs_t_flat} for which the autocorrelation function reaches the value $g^{(2)}(5T)\approx1.2\times10^{-4}$ at the end of the fifth mechanical period and stays within $\pm5\%$ of this value for $0.38$ mechanical periods.
All numerical simulations \yuxun{in \secname{discuss}} are also performed in a larger Hilbert space including up to $10$ photons and $25$ phonons so that the increasing excitation due to errors can be accurately calculated.

\subsection{Thermal Initial States}

Thus far, the mechanical initial state has been assumed to be the ground state. 
However, undesired thermal excitations in the mechanical oscillator are common in many experiments. 
It is therefore helpful to study a thermal initial state of the mechanical oscillator, which reads
\be
\hrho=\frac{1}{1+\bar{n}_m}\sum_{n=0}^\infty\left(\frac{\bar{n}_m}{1+\bar{n}_m} \right)^n\ket{n}\bra{n}\ ,
\ee
with $\bar{n}_m$ being the average phonon number initially in the mechanical mode.
Therefore, in this section, the error of the theoretical model is analyzed when the mechanical initial state is cooled to a near-ground state instead of a perfect ground state.

Through numerical simulation, it can be concluded that when the initial state is cooled to a near-ground state with the phonon number below the threshold $\bar{n}_m<0.1$, the value of the autocorrelation function at the end of the fifth mechanical period remains at a low value $g^{(2)}(5T)<1.3\times10^{-4}$.
However, for thermal states with higher phonon population $\bar{n}_m=10$, the autocorrelation function $g^{(2)}(5T)$ increases by an order of magnitude to $3.7\times10^{-3}$ which suggests a significant decrease in the strength of the photon blockade.
This decrease can be explained by the larger uncertainty in the position of the\raw{ thermal} oscillator \yuxun{due to the nature of the thermal state}.
When this uncertainty becomes comparable to the displacement of the oscillator induced by the optical field, there is a non-vanishing probability for another photon to enter the cavity and thus the strength of the photon blockade effect is diminished.

Although it is not a trivial task to perform a near-ground-state cooling, there are extensive research works~\cite{YWB+19,NM21,BKU+22} on this topic with less-than-unity average phonon number $\bar{n}_b$.

\subsection{Mechanical Dissipation}

In Sec. \ref{sec:num}, we assumed that the interactions between the mechanical oscillator and the environment are negligible.
In any realistic setting, however, mechanical noise is likely to play an important role. Here, we examine the impact of thermal noise and dissipation on our results.
For optomechanical systems, mechanical dissipation at a rate $\gamma$ in an environment with mean thermal occupation $\bar{n}_b$ can be modelled by the Lindbladian $\mathat{L}_m+\mathat{L}_p$~\cite{HHL+15}.
The term $\hat{\mathcal{L}}_m$ describes the thermalization of the mechanical mode in a frame displaced by the optical field and reads
\be
\hat{\mathcal{L}}_m=&\gamma(\bar{n}_b+1)\mathat{D}[\hb-\frac{g_0}{\omega_m}\ha^\dag \ha]\\
&+\gamma\bar{n}_b\mathat{D}[\hb^\dag-\frac{g_0}{\omega_m}\ha^\dag \ha]\ ,
\ee
where the superoperator $\mathat{D}[\hat{o}]$ satisfying the equation
\be
\mathat{D}[\hat{o}]\hrho=\hat{o}\hrho \hat{o}^\dag-\frac{1}{2}(\hat{o}^\dag \hat{o}\hrho+\hrho \hat{o}^\dag \hat{o})
\ee
is the Lindblad superoperator.
Due to the optomechanical coupling, there is also an extra optical dephasing term
\be
\mathat{L}_p=\frac{4\gamma g_0^2}{\omega_m^2\ln(1+1/\bar{n}_b)}\mathat{D}[\ha^\dag \ha]
\ee
induced by the mechanical dissipation.

\begin{figure}[t]
   \centering
   \includegraphics[keepaspectratio, width=0.45\textwidth]{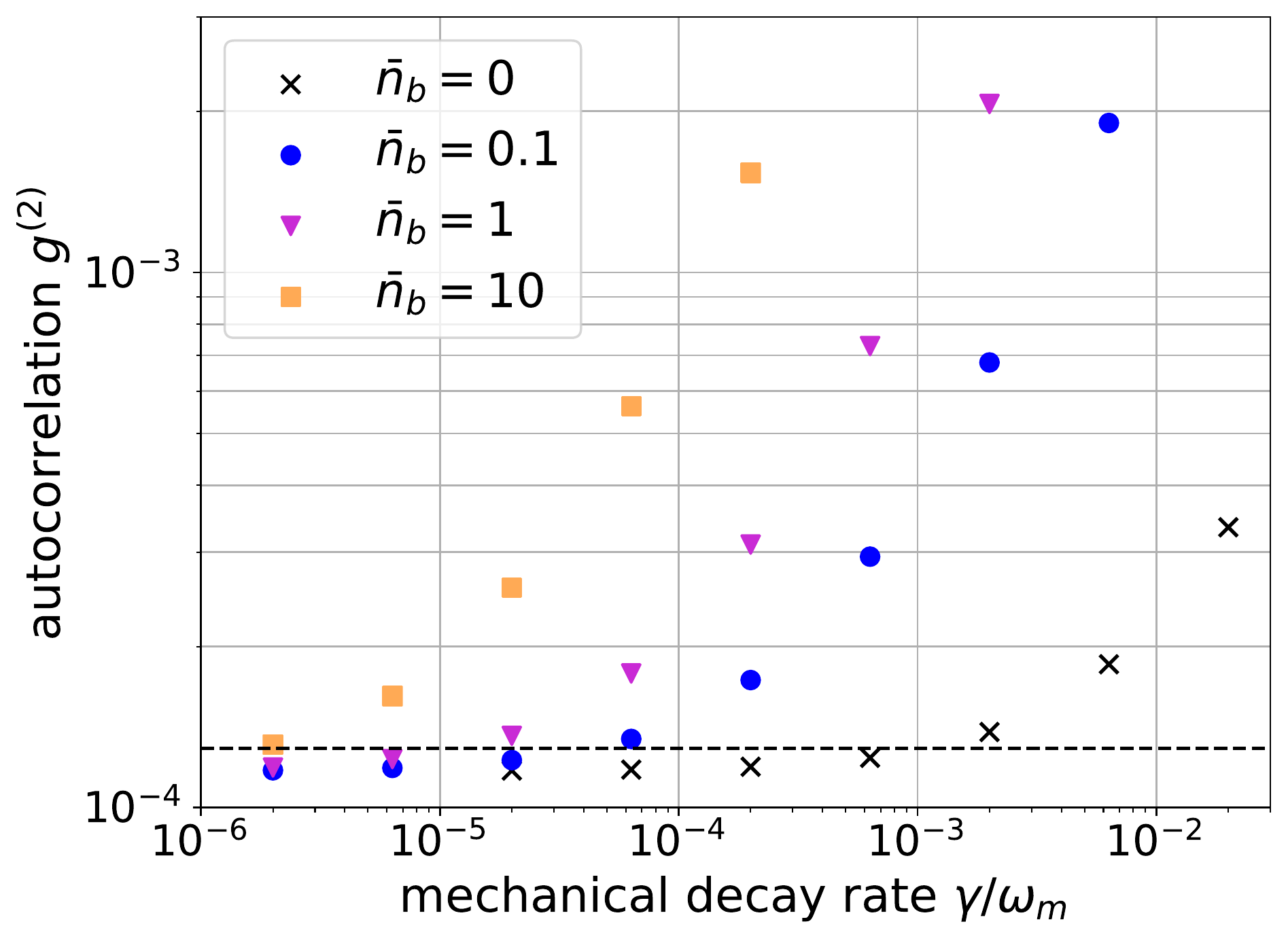}
   \caption{The autocorrelation function $g^{(2)}$ as a function of the relative mechanical decay rate $\gamma/\omega_m$ for average phonon number in the heat bath being $\bar{n}_b=0$ (cross), $\bar{n}_b=0.1$ (circle), $\bar{n}_b=1$ (triangle), and $\bar{n}_b=10$ (square). The dashed line specifies the value of the autocorrelation function that is $10\%$ larger than the value without the mechanical loss (i.e. $\gamma=0$). The assumptions of vanishing mechanical decay rate $\gamma\ll\omega_m$ and environmental phonon occupation $\bar{n_b}\approx0$ are required in order to achieve strong photon blockade.}
   \label{fig:mech_diss} 
 \end{figure}

\yuxun{To determine the effect of mechanical noise on the photon blockade, we plot the autocorrelation function $g^{(2)}(5T)$  numerically for finite mechanical loss rate $\gamma$ for different values of the environmental thermal populations (i.e. $\bar{n}_b=0,0.1,1,10$).
\Figname{mech_diss} shows the change in the autocorrelation function at the end of the fifth mechanical period as a function of the mechanical decay $\gamma/\omega_m$ in units rescaled by the mechanical frequency. The values of $\gamma$ range from the negligible value $\gamma=2\times10^{-6}\omega_m$ to  $\gamma=2\times10^{-2}\omega_m$ which equals to the optical decay $\kappa$.

From \figname{mech_diss}, we see that a higher dissipation rate $\gamma$ leads to higher values of autocorrelation $g^{(2)}$, which implies a weaker photon blockade.
In a zero-temperature environment (i.e. $\bar{n}_b=0$, indicated by the crosses), the autocorrelation function appears robust to noise (i.e., it stays at the same order of magnitude $g^{(2)}\sim10^{-4}$) even when $\gamma$ becomes comparable to the cavity decay $\kappa = 2\times10^{-2}\omega_m$.}
This suggests that, in the zero-temperature environment, the standard of negligible mechanical decay can be met by a variety of experimental setups~\cite{AKM14,KPJ+11,MZO+16,RMM+20}.
On the other hand, for environmental thermal occupation larger than the value $\bar{n}_b=1$, an extremely low mechanical decay rate $\gamma\approx10^{-6}\omega_m$ is required for the autocorrelation function to deviate by less than $10\%$ from its original value.
Furthermore, the autocorrelation function increases by more than one order of magnitude when the mechanical decay rate $\gamma$ becomes comparable to the optical decay rate $\kappa$.
Although an autocorrelation of value $\sim10^{-3}$ does not meet the criterion of strong photon blockade used in this work, it is still more than one order of magnitude smaller than the best value that can be achieved with a monochromatic driving as in \secname{num_mono}.

The minimum value of the autocorrelation function $g^{(2)}$ has a super-linear dependence on both the mean phonon occupation $\bar{n}_b$ in the heat bath and the mechanical decay rate $\gamma$.
The increasing dependence on the temperature and the coupling rate of the thermal bath is expected \raw{since the system is driven by a weak optical field and thus a strongly coupled, hot bath can be regarded as a strong mechanical driving which dominates the system dynamics and renders the optimization invalid.}
\yuxun{because a strongly coupled, hot bath can create excitations in the cavity through optomechanical coupling, which dominates the system dynamics and renders the optimization invalid.}

\subsection{Validity of Perturbative Methods}\label{sec:modelInaccu}

\yuxun{The optimization performed on the approximate autocorrelation function $\tilde{g}^{(2)}$ (\eq{g20_smallHS}) is only valid as long as the perturbative solutions presented in Sec. \ref{sec:theoryMethods} hold. Their validity is based on the following two assumptions: 
the driving strength remains small (i.e. $(\epsilon/\omega_m)^2\ll1$), and the coupling strength remains small (i.e. $(g_0/\omega_m)^4\ll1$). Here, we investigate how the accuracy of $\tilde{g}^{(2)}$ changes as these two conditions break down.}

\begin{figure}[t]
   \centering
   \includegraphics[keepaspectratio, width=0.45\textwidth]{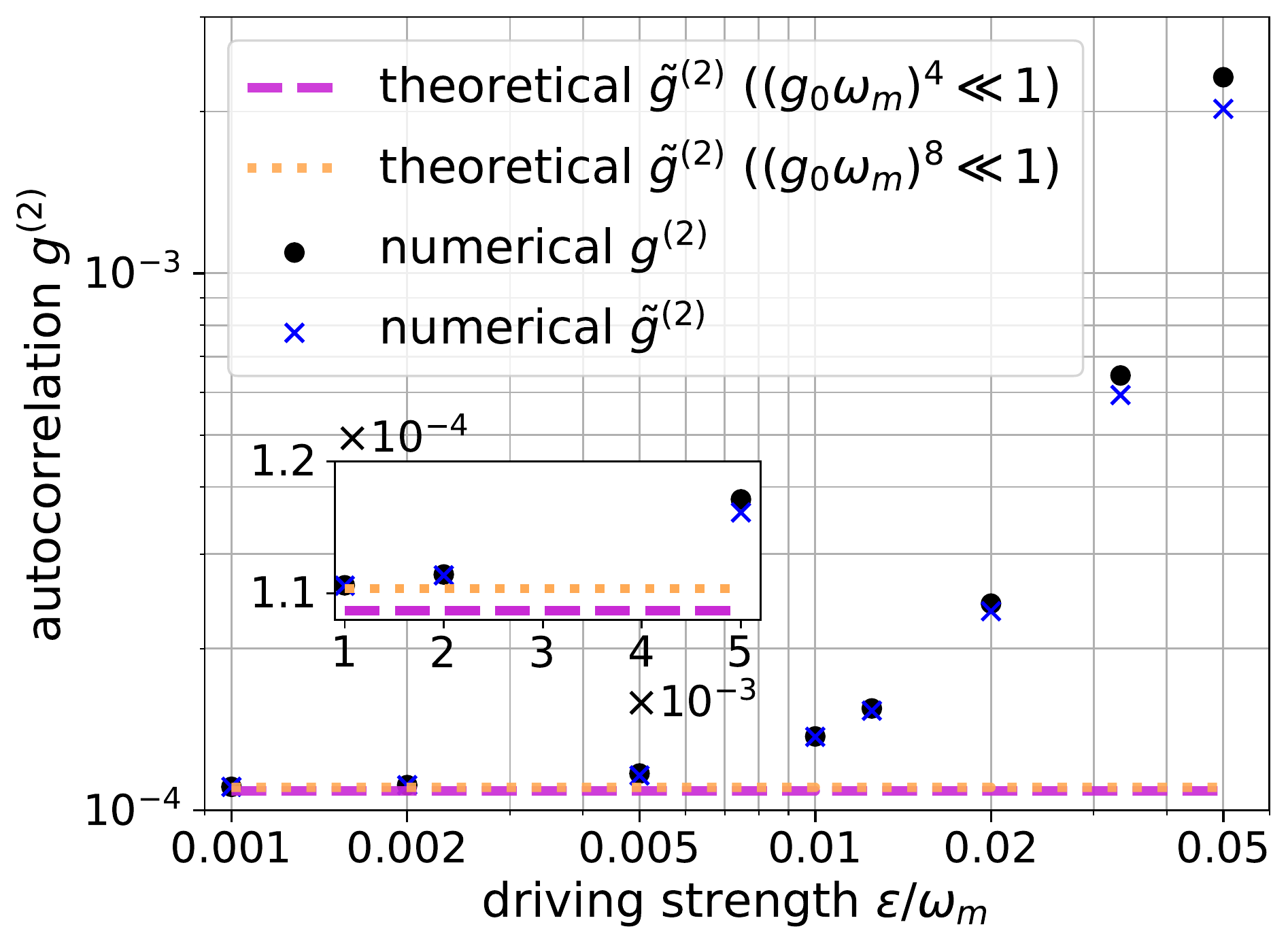}
   \caption{The autocorrelation $g^{(2)}$ and $\tilde{g}^{(2)}$ as functions of the driving strength $\epsilon/\omega_m$ rescaled by the mechanical frequency. 
The circular data points plot the exact autocorrelation values $g^{(2)}(5T)$ obtained from a full numerical simulation.
The cross data points plot the approximate autocorrelation values $\tilde{g}^{(2)}(5T)$ obtained from numerical simulation.
 The dashed curve plots the case when the perturbative expansion is truncated up to the order $(g_0/\omega_m)^2$ (i.e. $(g_0/\omega_m)^4\ll1$).
The dotted curve plots the case when the relevant perturbative expansion is truncated to a higher order which imposes a less strict condition (i.e. $(g_0/\omega_m)^8\ll1$) on the coupling strength.
The inset zooms into the results for small driving strengths.
The figure shows that a weak driving strength is necessary for obtaining the results in this work.}
   \label{fig:theory_error} 
 \end{figure}

The points at which the two perturbative assumptions break down can be found by studying the plot in \figname{theory_error} where the values of the autocorrelation function $g^{(2)}(5T)$ and the approximate autocorrelation function $\tilde{g}^{(2)}(5T)$ are plotted as functions of the driving strength.
The coupling strength $g_0$ is kept fixed throughout\footnote{This is because a change in the coupling strength $g_0$ requires a re-optimization of the driving parameters, and the balance between the strength of the autocorrelation function and the width of the local minimum can be different for each local minimum of the target function $\eta_1$. 
Therefore, instead of varying the coupling strength $g_0$, the perturbative expansion that assumes a small coupling strength is truncated to a higher order such that only the assumption $(g_0/\omega_m)^{12}\ll1$ is required to hold. }
The dashed and dotted lines show $\tilde{g}^{(2)}$ when terms up to orders $(g_0/\omega_m)^2$ and $(g_0/\omega_m)^6$ are included, respectively.  
The circular dots and the crosses show numerical simulations of $g^{(2)}$ and $\tilde{g}^{(2)}$, respectively. 

Firstly, it can be observed that the theoretical values for $\tilde{g}^{(2)}(5T)$ (dashed and dotted lines)  stay constant in \figname{theory_error}.
This can be explained by the following argument.
Because the approximate autocorrelation function $\tilde{g}^{(2)}$ omits the contribution from the three and more-photon populations, and because the two-photon population is negligible near its minimum, the function $\tilde{g}^{(2)}$ is approximately proportional to the ratio $p_2/p_1^2$.
According to the Neumann series in \eq{Neumann}, the single-photon occupation is dominated by the second-order term $\mathat{S}^{(2)}_1$ which is proportional to $\epsilon^{2}$. The two-photon occupation is dominated by the fourth-order term $\mathat{S}^{(4)}_1$, which is proportional to $\epsilon^4$. 
Thus, to leading order, the approximate autocorrelation function is independent of the driving strength. As a result, the magnitude of $\epsilon$ does not influence $g^{(2)}$.
In contrast, the three-photon occupation is included in the full numerical calculation of the autocorrelation function $g^{(2)}$, and thus its value changes as a function of the driving strength.

Next, the numerically obtained values for $g^{(2)}$ and $\tilde{g}^{(2)}$ (circles and crosses) can be compared with the theoretical values (lines). 
For a very small driving strength $\epsilon=\omega_m/1000$ (\figname{theory_error} inset), the majority of the error around the minimum can be attributed to the assumption of small coupling strength ($(g_0/\omega_m)^4\ll1$) as there is a wider disagreement between the results when the relevant perturbative expansion is truncated to different orders (dashed and dotted lines) than between the theoretical values and the numerical values (lines and scattered data).
However, the overall error of the theoretical model is negligible ($<0.01$) even when the terms quartic in the coupling strength are truncated. 
On the other hand, for stronger driving in \figname{theory_error}, the error is dominated by inaccuracy due to the truncated Hilbert space in the theoretical model.
Both higher-order correction to the single-photon and two-photon populations (i.e. $p_1$ and $p_2$) and the three-photon population can explain the diversion of the exact autocorrelation values (circular dots) from the theoretical values (lines).
However, the relatively minor difference between the exact values of the autocorrelation function $g^{(2)}$ and the numerically simulated values of the approximate autocorrelation function $\tilde{g}^{(2)}$ suggests that the higher order corrections to the single- and two-photon populations are responsible for the majority of the error.
In the presence of these errors, the optimized driving parameters may not lead to the true minimum of the autocorrelation function, but still, lead to a fast decrease in the value of the autocorrelation function near the specified time $t=5T$.

Overall, according to \figname{theory_error}, it is safe to conclude that the analytic approximations only start breaking down for values larger than $\epsilon/\omega_m \approx 0.005$, which is the value used in Sec. \ref{sec:num}.

\section{Conclusions and Outlook}

While it is elusive to realize a strong photon blockade effect without the clear separation of scales, the present control scheme can substantially improve the fidelity of photon-blockaded quantum states under imperfect conditions.
Since the question of what can be practically achieved with suitably shaped driving depends on the actual system parameters, a brief overview over system parameters that have been realized in existing optomechanical platforms is in order.
Table~\ref{tbl:experiments} shows the relevant parameters from a selection of works, together with the calculated autocorrelation function. 
The autocorrelation values are computed by minimizing the target function $\tilde{g}^{(2)}(5T)$ at target time $t_{op}=5T$ with respect to the driving parameters shown for each case.
For each of the examples, two different temperatures (i.e. $\bar{n}_b=0,1$) of the thermal bath surrounding the mechanical oscillator are shown.

\begin{table*}[t]
\begin{tabular}{|l|l|l|l|l|l|l|l|}
\hline
                                                      & $g_0/\omega_m$     & $\kappa/\omega_m$  & $\gamma/\omega_m$   & $g_0/\kappa$  & $g^{(2)}(5T)$ ($\bar{n}_b=0$) & $g^{(2)}(5T)$ ($\bar{n}_b=1$)     \\ \hline
Nano-acoustic resonator~\cite{MRL+20} & $1.4\times10^{-4}$ & $2.3\times10^{-1}$ & $4.2\times10^{-11}$ & $6.3\times10^{-4}$ & $1.0$ & $1.0$ \\ \hline
Hybrid system (theory)~\cite{DGH19}   & $4.0\times10^{-1}$ & $4.6\times10^{-2}$ & $1.0\times10^{-2}$  & $8.7$   & $1.5\times10^{-3}$ &  $1.5\times10^{-2}$       \\ \hline
Parameters used in this work                          & $3.0\times10^{-1}$  & $2.0\times10^{-2}$ & $0$                 & $15$       & $5.8\times10^{-5}$ & $5.8\times10^{-5}$        \\ \hline
\end{tabular}
\caption{System parameters in experiments in comparison with the parameters adopted in this work and the strength of the resulting photon blockade at the target time $t_{op}=5T$.
The coupling strength assumed in this work is too large in comparison with experimental data in the work on nano-acoustic resonators.
The parameters in the theoretical work~\cite{DGH19} suggest that hybrid systems potentially constitute good platforms for realizing photon blockade.
}\label{tbl:experiments}
\end{table*}

While it is generally difficult to achieve strong photon blockade, many experimental proposals are moving towards the values assumed in this work.
Most difficulty originates from the requirement of strong coupling strength $g_0>\kappa$.
Especially for solid-state resonators, the best values~\cite{CAS+11,VDW+12,MRL+20,RMM+20} of the coupling strength are still three orders of magnitude smaller than the cavity linewidth.
As shown in Table~\ref{tbl:experiments}, the strength of photon blockade is negligible in such a case.
However, there are several theoretical proposals~\cite{XGD12,RBA+14,DGH19} for enhancing the optomechanical coupling strength.
For example, as listed in Table~\ref{tbl:experiments}, the coupling strength and the cavity decay rate in the hybrid optomechanical system~\cite{DGH19} are of the same order of magnitude as considered in this work. Thus, the hybrid system could potentially be a good platform to realize the results in \secname{num}.
Some optomechanical systems with levitated particles~\cite{MMG+08,SLZ+11,SJO+20,MMP+20} also have strong values of coupling strength $g_0$ as compared to the mechanical frequency $\omega_m$.
However, in such a system, it is usually true that the inequality $g_0>\omega_m$ holds, which is not consistent with the approximations used in the present approach.

Even though the present work thus does not directly apply to systems with \raw{levitated particles}strong coupling strength ($g_0> \omega_m$), one would expect that suitably designed bichromatic (or polychromatic) driving offers increased capabilities for the preparation of quantum mechanical states also in systems with stronger optomechanical interactions.
In particular, strong optomechanical interactions could be used for the preparation of broader classes of quantum states, and the driving techniques developed in this work can be generalized straightforwardly to any other quantum state with desirable properties. 
Stronger phonon-photon coupling would certainly make it generally more difficult to find driving patterns such that the system evolves towards the desired states.
However, the potential of the non-Gaussian quantum states that might be realized in such systems make such an endeavor a worthwhile goal.

\section{Acknowledgements}
\yuxun{We thank Markus Rademacher for helpful discussions. S.Q. is funded in part by the Wallenberg Initiative on Networks and Quantum Information (WINQ) and in part by the Marie Skłodowska--Curie Action IF programme “Nonlinear optomechanics for verification, utility, and sensing” (NOVUS) -- Grant-Number 101027183.}

\nocite{apsrev41Control}
\bibliographystyle{apsrev4-2}
\bibliography{refs}
\onecolumngrid

\newpage
\widetext

\appendix

\section{Theory}\label{apdx:theory}

In this Appendix, details on the solution to the Master equation in \eq{MasterEq} are presented.

\subsection{System dynamics}\label{apdx:sysDymSoln}

It is helpful to define the left-hand and the right-hand ladder superoperators $\hat{a}_L$, $\hat{b}_L$  and $\hat{a}_R$, $\hat{b}_R$ that satisfy the equations
\be
\hat{a}_L(\hrho)=\hat{a}\hrho,\qquad\hat{a}_R(\hrho)=\hrho\hat{a},\qquad\hat{b}_L(\hrho)=\hat{b}\hrho,\qquad\hat{b}_R(\hrho)=\hrho\hat{b}\ .
\ee

In terms of the ladder superoperators, the photon-conserving evolution operator $\mathat{S}_0$ defined in \secname{theoryMethods} reads
\be
\mathat{S}_0(t)=\mathat{S}_F(t)\mathat{S}_D(t)\mathat{S}_I(t)\mathat{S}_K(t)
\ee
with
\be
\mathat{S}_F(t)=&\exp\brkts{-i\omega_c(\ha^\dag_L \ha_L-\ha^\dag_R\ha_R)t}\exp\brkts{-i\omega_m(\hb^\dag_L \hb_L-\hb^\dag_R\hb_R)t}\ ,\\
\mathat{S}_D(t)=&\exp\brkts{-\frac{\kappa}{2}(\ha^\dag_L \ha_L+ \ha^\dag_R \ha_R)t}\ ,\\
\mathat{S}_I(t)=&\exp\brkts{f_1(t)(\ha_L^\dag \ha_L \hb_L^\dag-\ha_R^\dag \ha_R \hb_R)-h.c.}\ ,\\
\mathat{S}_K(t)=&\exp\brkts{if_2(t)\brkts{(\ha_L^\dag \ha_L)^2-(\ha_R^\dag \ha_R)^2}}\ .
\ee
The propagators $\mathat{S}_F(t)$, $\mathat{S}_D(t)$, $\mathat{S}_I(t)$, $\mathat{S}_K(t)$ correspond to the free evolution, the dephasing, the optomechanical interaction and the Kerr-type nonlinearity respectively and
\be
f_1(t)&=\frac{g_0}{\omega_m}(e^{i\omega_mt}-1)\ ,\\
f_2(t)&=\brkts{\frac{g_0}{\omega_m}}^2(\omega_mt-\sin(\omega_mt))\ ,
\ee
are real, time-dependent functions. 
In the coherent case without driving and dissipation (i.e. $\kappa=\xi(t)=0$), this solution coincides with results from earlier works~\cite{MMT97,BJK97}.

The photon-varying part $\mathat{L}_1(t)$ of the Master equation in the frame defined by the photon-conserving superoperator $\mathat{S}_0$ then reads
\be\label{eq:SvM}
\tilde{\mathcal M}(t)=\mathat{S}_0^{-1}(t)\mathat{L}_1(t)\mathat{S}_0(t)=-i(\xi^*\mathat{A}_L-\xi \mathat{A}_R+\xi\mathat{A}_L^\dag-\xi^* \mathat{A}_R^\dag)+\kappa \mathat{A}_L\mathat{A}_R\ ,
\ee
in which 
\be\label{eq:anniOpTrans}
\mathat{A}_L=& \ha_Le^{-(\kappa/2+i\omega_c) t}e^{if_2(2 \ha_L^\dag \ha_L-1)}e^{f_1  \hb^\dag_L-f_1^* \hb_L}\ ,\\
 \mathat{A}_L^\dag= & \ha_L^\dag e^{(\kappa/2+i\omega_c) t}e^{-if_2(2 \ha_L^\dag \ha_L+1)}e^{f_1^* \hb_L-f_1  \hb^\dag_L}\ ,\\
 \mathat{A}_R=& \ha_Re^{(-\kappa/2+i\omega_c) t}e^{-if_2(2 \ha_R^\dag \ha_R-1)}e^{f_1^*  \hb^\dag_R-f_1 \hb_R}\ ,\\
 \mathat{A}_R^\dag= & \ha_R^\dag e^{(\kappa/2-i\omega_c) t}e^{if_2(2 \ha_R^\dag \ha_R+1)}e^{f_1  \hb_R-f_1^* \hb_R^\dag}\\
\ee
are the creation and annihilation superoperators in the frame defined by the superoperator $\mathat{S}_0$.
%The Neumann expansion (\eq{Neumann}) of $\mathat{S}_1$ cannot be expressed in terms of elementary functions because the integration of $\mathat{A}$ is not analytically approachable.
Given any initial state $\hrho_i$, the final state $\hrho_f$ can thus be perturbatively expanded using the Neumann series (\eq{Neumann}) into the form of the following sum of integrals
\be\label{eq:finSInt}
\hrho_f=\sum_{k=0}^\infty\int_{\bm{t^{(k)}}\in\mathbb{T}}\brkts{\mathat{S}_0(t)\prod_{j=1}^k\tilde{\mathcal{M}}(t_j)\hrho_i} d\bm{t^{(k)}}\ ,
\ee
with $\tilde{\mathcal{M}}$ defined in \eq{SvM} and the vector $\bm{t^{(k)}}$ of time-ordered variables $t_k<\hdots<t_1$, and the hyperpyramidal domain $\mathbb{T}$ being $0<t_k<\hdots<t_1<t$ as defined in \eq{Neumann}.
This can be taken as a starting point for the calculation of expectation values.

\subsection{Dynamics of The Vacuum State}\label{apdx:dymVacState}

One natural choice of the initial state is the vacuum state $\hrho_i=\ketbra{0}_c\otimes\ketbra{0}_m$.
Whenever the Neumann series in \eq{finSInt} is truncated to finite order, the final state of the system given a vacuum initial state is a finite superposition
\be\label{eq:finSGeneral}
\hrho_f(t)=\sum_{j=0}^Ng_j(t)\hrho(n_{L,j},n_{R,j},\beta_{L,j},\beta_{R,j})\ ,
\ee
with $N$ being a finite integer, $g_j$ being time-dependent, scalar functions and the density matrices 
\be
\hrho(n_L,n_R,\beta_L,\beta_R)=\ketbra{n_L}{n_R}_c\otimes \ketbra{\beta_L}{\beta_R}_m\ ,
\ee 
with the optical components $\ketbra{n_L}{n_R}_c$ in the Fock basis, and the mechanical states $\ketbra{\beta_L}{\beta_R}_m$ in the coherent basis.
As a result, the photon statistics of the final state can be extracted from corresponding components that satisfy the relation $n_L=n_R$.

To prove that the final state can be written as a finite sum of matrices $\hrho(n_L,n_R,\beta_L,\beta_R)$, it suffices to show that any matrix in the form $\hrho(n_L,n_R,\beta_L,\beta_R)$ is mapped to another matrix in the same form when acted on by the superoperators  $\mathat{A}_L^{(\dag)}$, $\mathat{A}_R^{(\dag)}$, and $\mathat{S}_0$.
This is because the expression $\mathat{S}_0(t)\prod\tilde{\mathcal{M}}(t_j)$ in each of the $k$-th order Neumann term (\eq{finSInt}) is a finite polynomial of the superoperators $\mathat{A}_L^{(\dag)}$, $\mathat{A}_R^{(\dag)}$, and $\mathat{S}_0$.

The action of the superoperators on the matrix $\hrho(n_L,n_R,\beta_L,\beta_R)$ reads
\be
\mathat{A}_L\hrho(n_L,n_R,\beta_L,\beta_R)&=g_L\hrho(n_L-1,n_R,\beta_L+f_1,\beta_R)\ ,\\
\mathat{A}_R\hrho(n_L,n_R,\beta_L,\beta_R)&=g_R\hrho(n_L,n_R-1,\beta_L,\beta_R+f_1^*)\ ,\\
\mathat{S}_0\hrho(n_L,n_R,\beta_L,\beta_R)&=g_c\hrho(n_L,n_R,e^{-i\omega_mt}(\beta_L+f_1n_L),e^{i\omega_mt}(\beta_R+f^*_1n_R))
\ee
with
\be
g_L&=\sqrt{n_L}e^{-(\kappa/2+i\omega_c) t}e^{if_2(2 n_L-1)}e^{(f_1  \beta_L^*-f_1^* \beta_L)/2}\ ,\\
g_R&=\sqrt{n_R}e^{(-\kappa/2+i\omega_c) t}e^{-if_2(2 n_R-1)}e^{(f_1^*  \beta_R^*-f_1 \beta_R)/2}\ ,\\
g_c&=e^{-\kappa(n_L+n_R)t/2}e^{-i\omega_c(n_L-n_R)t}e^{n_L(f_1\beta_L^*-h.c.)/2}e^{n_R(f_1^*\beta_R^*-h.c.)/2}e^{f_2(n_L^2-n_R^2)}
\ee
being scalar, time-dependent functions, and thus \eq{finSGeneral} is proven.
In the next section, the photon statistics obtained by partial tracing out the mechanical degree of freedom of the matrix elements satisfying the relation $n_L=n_R$ is presented.

\subsection{Photon Occupation}\label{apdx:finRes_theory}

Since the approximate autocorrelation $\tilde{g}^{(2)}$ in \eq{g20_smallHS} is a function of photon occupation $p_1$ and $p_2$ only, it is sufficient to calculate only the corresponding matrix elements in the density matrix in \eq{finSGeneral}.

In the regime of weak driving ($\abs{\xi}^2\ll\omega_m^2$), the leading order single-photon occupation $p_1(t)$ at time $t$ appears in the second order of the perturbative series in \eq{finSInt} and reads
\be\label{eq:1photOccuRaw}
p_1(t)=2\Re\int_0^t\int_0^{t_1}\exp\brkts{\frac{g_0^2}{\omega_m^2}(e^{i\omega_m(t_1-t_2)}-1)+i\frac{g_0^2}{\omega_m}(t_2-t_1)-\kappa t+\frac{\kappa}{2}(t_1+t_2)}\zeta(t_1)\zeta^*(t_2)dt_2dt_1\ ,
\ee
where $\Re$ denotes the real part of the expression and $\zeta(t)=\xi(t)\exp(i\omega_ct)$ being the driving function in the frame rotating at the cavity resonance frequency.

The leading order contribution to two-photon occupation $p_2(t)$ at time $t$ appears in the fourth order of the series in \eq{finSInt} (i.e. $\propto \abs{\xi}^4$) and can be written as
\be\label{eq:2photOccuRaw}
p_2(t)=4\Re\int_{\bm{t^{(4)}}\in\mathbb{T}}\exp\brkts{-\frac{2g_0^2}{\omega_m^2}+\frac{\kappa}{2}(t_1+t_2+t_3+t_4-4t)}\sum_{j=1}^{3}h_j(t,\bm{t^{(4)}})d\bm{t^{(4)}}\ ,
\ee
with the vector $\bm{t^{(4)}}=(t_1,t_2,t_3,t_4)$ of time-ordered variables and the domain $\mathbb{T}$ as defined in \eq{Neumann}, and the time-dependent functions $h_j(t,\bm{t^{(4)}})$ defined as
\be\label{eq:2photOccuComponents}
h_1(t,\bm{t^{(4)}})=&\exp\brkts{\frac{g_0^2}{\omega_m^2}\sum_{1\le j<k\le4}s^{(1,1)}_{j,k}\exp\brkts{s^{(2,1)}_{j,k}(t_j-t_k)}-s^{(1,1)}_{j,k}s^{(2,1)}_{j,k}(t_j-t_k)}\zeta^*(t_1)\zeta^*(t_2)\zeta(t_3)\zeta(t_4)\ ,\\
h_2(t,\bm{t^{(4)}})=&\exp\brkts{\frac{g_0^2}{\omega_m^2}\sum_{1\le j<k\le4}s^{(1,2)}_{j,k}\exp\brkts{s^{(2,2)}_{j,k}(t_j-t_k)}-s^{(1,2)}_{j,k}s^{(2,2)}_{j,k}(t_j-t_k)}\zeta^*(t_1)\zeta(t_2)\zeta^*(t_3)\zeta(t_4)\ ,\\
h_3(t,\bm{t^{(4)}})=&\exp\brkts{\frac{g_0^2}{\omega_m^2}\sum_{1\le j<k\le4}s^{(1,3)}_{j,k}\exp\brkts{s^{(2,3)}_{j,k}(t_j-t_k)}-s^{(1,3)}_{j,k}s^{(2,3)}_{j,k}(t_j-t_k)}\zeta(t_1)\zeta^*(t_2)\zeta^*(t_3)\zeta(t_4)\ ,\\
\ee
and the expressions $s^{(1,i)}_{j,k}$ and $s^{(2,i)}_{j,k}$ are either positive or minus sign as defined in Table~\ref{tbl:slijk}.

\begin{table}[htbp]
\begin{tabular}{|c|c|c|c|c|c|c|c|}
\hline
%\diagbox{$(l,i)$}{$(j,k)$}   	& $(1,2)$     	  	& $(1,3)$  		& $(1,4)$   			& $(2,3)$  			& $(2,4)$  		& $(3,4)$     \\ \hline
%$(1,1)$ 					& $-$ 			& $+$ 			& $+$ 				& $+$ 				& $+$ 			& $-$ 		\\ \hline
%$(1,2)$ 					& $+$	  	  		& $-$ 		     	& $+$			       	& $+$			 	 	& $-$ 			& $+$    		\\ \hline
%$(1,3)$ 					& $+$ 			& $+$ 			& $-$				  	& $-$  		 		& $+$			 	& $+$       	\\ \hline
%$(2,1)$ 					& $+$ 			& $-$ 			& $-$ 				& $-$ 				& $-$ 			& $-$ 	\\ \hline
%$(2,2)$ 					& $-$    			& $+$      		& $+$       			& $-$  				& $-$ 			& $-$    	\\ \hline
%$(2,3)$ 					& $+$				& $+$  			& $+$   				& $-$				 	& $-$       		& $-$		\\ \hline
 \diagbox{$(j,k)$}{$(l,i)$} &$(1,1)$ &$(1,2)$ &$(1,3)$ &$(2,1)$ &$(2,2)$ &$(2,3)$ \\ \hline
 $(1,2)$ & $-$ & $+$ & $+$ & $+$ & $-$ & $+$ \\ \hline
 $(1,3)$ & $+$ & $-$ & $+$ & $-$ & $+$ & $+$ \\ \hline
 $(1,4)$ & $+$ & $+$ & $-$ & $-$ & $+$ & $+$ \\ \hline
 $(2,3)$ & $+$ & $+$ & $-$ & $-$ & $-$ & $-$ \\ \hline
 $(2,4)$ & $+$ & $-$ & $+$ & $-$ & $-$ & $-$ \\ \hline
 $(3,4)$ & $-$ & $+$ & $+$ & $-$ & $-$ & $-$ \\ \hline
\end{tabular}
\caption{The values of the sign variables $s^{(l,i)}_{j,k}$ in \eq{2photOccuComponents}.}\label{tbl:slijk}
\end{table}

%\end{widetext}

The integral in \eq{2photOccuRaw} contains double exponentials in the form of $\exp(\exp(ix))$ which cannot be integrated into a closed form and numerical tools are inefficient for multiple integrals.
However, given a small coupling strength that satisfies the inequality $g^4_0/\omega_m^4\ll1$, the double exponential can be expanded into the following form
\be\label{eq:Taylor}
\exp(\frac{g_0^2}{\omega_m^2}e^{i\omega_m(t_p-t_q)})\approx1+\frac{g_0^{2}e^{i\omega_m(t_p-t_q)}}{\omega_m^{2}}\ ,
\ee
using the Taylor expansion.
The right-hand side of \eq{Taylor} can be integrated exactly but the results are too long to be presented in the text.
Interested readers can find the full results online~\footnote{Fast-Photon-Blockade-Codes (Version 1.0.0) [Github Repository]. github.com/lingyx97/Fast-Photon-Blockade-Codes}.

\section{Parameters Used for Figures}\label{apdx:params}
In this section, the optimized parameters used for figures in the main text are listed.
All parameters are normalized in units of the mechanical frequency (i.e. $\omega_m=1$).
For example, one unit of time is one mechanical period, and one unit frequency is the mechanical resonance frequency.

\Figname{mono_g2_vs_t} fixes the driving strength $\epsilon=1/200$, and the cavity decay rate $\kappa=1/50$.
The evolution time at which the autocorrelation function is minimized $t_{op}$, the coupling strength $g_0$ and the optimized driving detuning $\Delta$ are listed in Table~\ref{tbl:fig1params}.
\begin{table}[htbp]
\begin{tabular}{||c|c|c|| c|c|c|| c|c|c|| c|c|c|| c|c|c||}
\hline
$\bm{t_{op}/T}$ & $\bm{g_0}$ & $\bm\Delta$ & $\bm{t_{op}/T}$ & $\bm{g_0}$ & $\bm\Delta$ & $\bm{t_{op}/T}$ & $\bm{g_0}$ & $\bm\Delta$ 	& $\bm{t_{op}/T}$ 	& $\bm{g_0}$	& $\bm \Delta$ 	& $\bm{t_{op}/T}$ 	& $\bm{g_0}$ 	& $\bm\Delta$ \\	\hhline{||=|=|=|| =|=|=|| =|=|=|| =|=|=|| =|=|=||}
3   & 0.3   & 0.055491 & 3   & 0.35  & -0.01397 & 3   & 0.4   & -0.09137	& 3   	& 0.5  	& -0.22977 	& 3   	& 0.6   	& -0.41061	\\ \hline
4   & 0.3   & -0.00498 & 4   & 0.35  & -0.07028 & 4   & 0.4   & -0.1376	& 4   	& 0.5  	& -0.22298 	& 4   	& 0.6   	& -0.38487	 \\ \hline
5   & 0.3   & -0.03953 & 5   & 0.35  & -0.09842 & 5   & 0.4   & -0.0986 	& 5   	& 0.5  	& -0.24291 	& 5   	& 0.6   	& -0.37614   	\\ \hline
6   & 0.3   & -0.06016 & 6   & 0.35  & -0.06547 & 6   & 0.4   & -0.1269 	& 6   	& 0.5  	& -0.23397 	& 6   	& 0.6   	& -0.38582	 \\ \hline
7   & 0.3   & -0.06974 & 7   & 0.35  & -0.08281 & 7   & 0.4   & -0.14577	& 7   	& 0.5  	& -0.24525 	& 7   	& 0.6   	& -0.37771	 \\ \hline
8   & 0.3   & -0.04821 & 8   & 0.35  & -0.09852 & 8   & 0.4   & -0.13611 	& 8   	& 0.5  	& -0.23984 	& 8   	& 0.6   	& -0.36937	\\ \hline
9   & 0.3   & -0.05612 & 9   & 0.35  & -0.10887 & 9   & 0.4   & -0.14063	& 9   	& 0.5  	& -0.24643 	& 9   	& 0.6   	& -0.36882	 \\ \hline
10  & 0.3   & -0.06519 & 10  & 0.35  & -0.10828 & 10  & 0.4   & -0.15001 	& 10   	& 0.5  	& -0.24317 	& 10   	& 0.6   	& -0.37111	\\ \hline
11  & 0.3   & -0.07255 & 11  & 0.35  & -0.10351 & 11  & 0.4   & -0.15037 	& 11   	& 0.5  	& -0.24718 	& 11  	& 0.6   	& -0.36774	\\ \hline
12  & 0.3   & -0.07763 & 12  & 0.35  & -0.10879 & 12  & 0.4   & -0.14792	& 12   	& 0.5  	& -0.24516 	& 12   	& 0.6   	& -0.36490	 \\ \hline
13  & 0.3   & -0.07932 & 13  & 0.35  & -0.11361 & 13  & 0.4   & -0.15261 	& 13   	& 0.5  	& -0.24770 	& 13   	& 0.6   	& -0.36573	\\ \hline
14  & 0.3   & -0.07634 & 14  & 0.35  & -0.11482 & 14  & 0.4   & -0.15415	& 14   	& 0.5  	& -0.24642 	& 14   	& 0.6   	& -0.36584	 \\ \hline
15  & 0.3   & -0.07642 & 15  & 0.35  & -0.11248 & 15  & 0.4   & -0.15218 	& 15   	& 0.5  	& -0.24808 	& 15  	& 0.6   	& -0.36412	\\ \hline
\end{tabular}
\caption{Optimized parameters for \figname{mono_g2_vs_t}.}\label{tbl:fig1params}
\end{table}\FloatBarrier

\Figname{bichrom_raw}  fixes the driving strength $\epsilon=\epsilon_1=\epsilon_2=1/200$, the cavity decay rate $\kappa=1/50$, the coupling strength $g_0=0.3$ and the evolution time at which the autocorrelation function is minimized $t_{op}=5T$.
The optimized monochromatic detuning reads $\Delta=-0.03953$.
The optimized bichromatic detunings read $\Delta_1=-0.0192947$, $\Delta_2=-0.0282437$, and the phase difference between the two driving fields reads $\psi=2.83667$.

\Figname{bi_g2_vs_kap} fixes the driving strength $\epsilon_1=\epsilon_2=1/200$, the coupling strength $g_0=0.3$, the evolution time $t_{op}=5T$.
The cavity decay $\kappa$, the optimized driving detunings $\Delta_1$, $\Delta_2$, and the optimized phase difference between the two driving $\psi$ are listed in Table~\ref{tbl:fig3params}.
\begin{table}[htbp]
\begin{tabular}{|c||c|c|c|c|c|c|c|c|c|c|}
\hline
$\bm{t_{op}/T}$ & 5 & 5 & 5 & 5 & 5 & 5 & 5 & 5 & 5 & 5 \\ \hline
$\bm\kappa$ & $0.01$ & $0.02$ & $0.03$ & $0.04$ & $0.05$ & $0.06$ & $0.07$ & $0.08$ & $0.09$ & $0.1$ \\ \hline
$\bm{\Delta_1}$ & $-0.03492$ & $-0.01929$ & $-0.01616$ & $-0.0092$  & $-0.00843$ & $-0.00877$ & $-0.00756$ & $-0.00822$ & $-0.00804$ & $-0.00752$ \\ \hline
$\bm{\Delta_2}$ & $-0.02443$ & $-0.02824$ & $-0.01193$ & $-0.00477$ & $-0.00496$ & $-0.00433$ & $-0.00465$ & $-0.00389$ & $-0.00375$ & $-0.00379$ \\ \hline
$\bm\psi$     & $3.51878$  & $2.83667 $ & $3.27736$  & $3.27906$  & $3.24947$  & $3.27937$  & $3.23194$  & $3.27625$  & $3.27532$  & 3.25781  \\ \hline
 \hline
$\bm{t_{op}/T}$ & 10 & 10 & 10 & 10 & 10 & 10 & 10 & 10 & 10 & 10 \\ \hline
$\bm\kappa$ & 0.01 & 0.02 & 0.03 & 0.04 & 0.05 & 0.06 & 0.07 & 0.08 & 0.09 & 0.1 \\ \hline
$\bm{\Delta_1}$ & -0.04765 & -0.11406 & -0.03877 & -0.04014  &-0.04077 & -0.04413 & -0.04394 & 0.95322 &-0.04514 & -0.04499 \\ \hline
$\bm{\Delta_2}$ & -0.05129 & -0.11768 & -0.04053 & -0.04133 & -0.04378 & -0.04292 & -0.04459 & 0.95523 & -0.04852 & -0.04600 \\ \hline
$\bm\psi$     & 2.89786  & 2.90915  & 3.032  & 3.06708  			& 2.95241  & 3.21805  & 3.10049  & 3.26766  & 2.92939  & 3.0779  \\ \hline
\end{tabular}
\caption{Optimized parameters for \figname{bi_g2_vs_kap}.}\label{tbl:fig3params}
\end{table}\FloatBarrier

\Figname{bi_g2_vs_t_flat} fixes the driving strength $\epsilon_1=\epsilon_2=1/200$, the cavity decay rate $\kappa=1/50$, the coupling strength $g_0=0.3$, and the evolution time at which the autocorrelation function is minimized $t_{op}=5$.
The optimized driving detunings are $\Delta_1=0.0165539$ and $\Delta_2=-0.0364786$, and the optimized phase difference is $\psi=1.4571$.

\end{document}